\documentclass{article}
\usepackage{graphicx} 
\usepackage[english]{babel}
\usepackage{amsthm}
\usepackage{amsmath}
\usepackage{amssymb}
\usepackage{enumitem}
\usepackage{xcolor}
\usepackage{comment}
\usepackage[sorting=none]{biblatex}
\usepackage{authblk}

\theoremstyle{plain}
\newtheorem{theorem}{Theorem}[section]
\newtheorem{proposition}[theorem]{Proposition}
\newtheorem{corollary}[theorem]{Corollary}
\newtheorem{lemma}[theorem]{Lemma}

\theoremstyle{definition}
\newtheorem{definition}[theorem]{Definition}

\theoremstyle{remark}
\newtheorem{remark}[theorem]{Remark}

\newcommand{\Pprob}{\mathbb P}

\title{Ferromagnetic Ising model on multiregular random graphs}

\author[1]{Diego Alberici}
\author[2]{Pierluigi Contucci}
\author[2]{Emanuele Mingione}
\author[2,3]{Filippo Zimmaro}

\affil[1]{\footnotesize\itshape DISIM, University of L'Aquila, L'Aquila, Italy}
\affil[2]{\footnotesize\itshape Department of Mathematics, University of Bologna, Bologna, Italy}
\affil[3]{\footnotesize\itshape Department of Computer Science, University of Pisa, Pisa, Italy}

\date{}

\begin{document}

\maketitle

\begin{abstract}
    A family of multispecies Ising models on generalized regular random graphs is investigated in the thermodynamic limit. The architecture is specified by class-dependent couplings and magnetic fields. We prove that the magnetizations, neighbours correlations and free energy converge to suitable functions evaluated at the solution of a belief propagation fixed point equation. In  absence of magnetic fields, a phase transition is identified and the corresponding critical parameters are determined by the spectral radius of a low-dimensional matrix.
\end{abstract}

\section{Introduction}
Statistical Mechanics on diluted networks provides remarkable examples of complex systems with fruitful applications \cite{mezard2009information,van2016random,van2024random}. The archetype is the ferromagnetic Ising model on the Erdos-R\'enyi random graph, or more general locally tree-like graphs \cite{leone2002ferromagnetic,desanctis,dembo2010ising,dembo2010gibbs,montanari2012weak,dommers2014ising}. Further examples cover the Potts and hard-core models \cite{dembo2013factor, dembo2014replica}, the antiferromagnetic Ising model \cite{contucci2013antiferromagnetic} and the monomer-dimer model \cite{zdeborova2006number,bls,alberici2014} on random graphs.
In certain circumstances, the Theoretical Physics ideas of cavity fields and belief progapagation \cite{mezard2009information} provide exact solutions and may be turned into rigorous proofs based on local weak convergence \cite{Aldous} and correlation inequalities.

The class of diluted Ising models we consider is characterized by an additional block (species) structure. More precisely, we assume that the $N$ particles of the system are partitioned in $n$ different species and we denote by $N_a$ the number of  particles of class $a$. Unlike in the standard mean-field case where the dilution is invariant with respect to the symmetric group acting on the $N$ particles, here such invariance is relaxed to the product of the symmetric groups acting on different blocks of particles. The large $N$ limit is taken keeping fixed the ratios $\alpha_a=N_a/N$. These models are often called multispecies models and  have been recently studied with both deterministic \cite{Gallo,Fedele, Lowe, Toth,Collet} and disordered interactions
\cite{MSK_original,panchenko_multi-SK, Partha,bates,baik2020,noi_deep2,Mourrat2020NonconvexII,
Chen2023on,DBMNL}.
The introduction of the multispecies structure has, roughly speaking, the effect of a dimensional growth of the degrees of freedom of the problem. 
For instance, in the multispecies version of the Curie Weiss model the solution is given by an $n$-dimensional variational principle and its stationary points represent the equilibrium values of the magnetization for each species \cite{Gallo}. We find that in diluted models, instead, the number of relevant parameters is smaller or equal to $n^2$. Compared with the mean-field case, this feature can be related to the presence of non trivial correlations in diluted models.

In this work we study the Ising model on a class of diluted random graphs that we call $\mathbf{k}$-regular random graphs, where $\mathbf{k}=(k_{ab})_{a,b\leq n}$ is a given, suitably chosen matrix. The value $k_{ab}$  represents the number of neighbours of class $b$ that a vertex of class $a$ has. The quenched randomness is the uniform measure on all the possible graph realizations of type $\mathbf{k}$ with given vertices. This class of random graphs is the multispecies generalization of standard $k$-regular random graphs, $k\in\mathbb{N}$, which correspond to the case $n=1$.
In the considered models, ferromagnetic couplings $\beta_{ab}>0$ and external fields $h_a$ are class-dependent.

We prove several results in the thermodynamic limit $N\to\infty$. First, for positive external fields, local magnetisations, neighbours' correlations and free energy are expressed as functions of the unique solution with positive entries of a multidimensional fixed point equation of dimension at most $n^2$. 
These results extend to non-negative external fields, at least one positive, under a general condition on the connectivity of the classes. Moreover, we prove that for arbitrary external fields and high temperature the local magnetisations converge to the unique solution of the fixed point equation. Finally, for zero external fields we show that the model undergoes a phase transition by computing its spontaneous magnetisation. The critical values of the parameters are identified: they are related to the spectral radius of a suitable matrix $\boldsymbol{M}$ (dimension at most $n^2\times n^2$), which comes from the linearisation of the recursion and encodes the connectivity of classes. 


Our proofs rely on the general ideas of cavity fields, belief propagation algorithms and local convergence, since the considered $\mathbf{k}$-regular random graphs are locally tree-like. Thus, we first focus on the existence and uniqueness of the solution of the fixed point equation obtained by the belief propagation recursion on the deterministic trees associated to our class of random graphs. Existence and uniqueness (in a suitable region) are proven by monotonicity and concavity arguments applied to the recursion. 
For non-negative fields, monotonicity and concavity follow from GKS and GHS correlation inequalities \cite{griffiths1967correlations, KS, griffiths1970concavity, lebowitz1974ghs}. However, aiming at going beyond positive fields, we provide a direct proof looking directly at the recursion: a suitable condition on the initialisation implies monotonicity and concavity without appealing to general correlation inequalities. This method allows us to identify three fixed points for zero external fields and low temperature and prove uniqueness of the fixed point for arbitrary fields and high temperature.
Eventually, in order to establish a connection between local observables on random graphs (local magnetisations and correlations) and the corresponding observables on trees, GKS and FKG correlation inequalities are used \cite{fortuin1971correlation}. The free energy is also computed, since its derivatives can be expressed in terms of local quantities on the graph.

As mentioned above, our approach allows to study the recursion beyond the case of non-negative fields. However, the lack of correlation inequalities does not always allow to move from trees to random graphs. 
In future works we aim at invesigating regimes with external fields of opposite signs and low temperature,  
due to the relevance of such models (e.g., Random Field Ising Models (RFIM) \cite{larkin1970effect,krzakala2010elusive}) and their applications in Physics \cite{klafter1984coulomb}, Chemistry \cite{de1984liquid}
and Social Sciences \cite{galam1997rational,zimmaro2024asymmetric}.




The paper is organised as follows. Section 2 presents the model and the main results. Sections 3 deals with the analysis of the recursion on trees. Section 4 covers the proofs of the main theorems by transposing the results of Section 3 from trees to the class of random graphs. A short Appendix collects auxilary results (part A) and recalls GKS, FKG, and GHS correlation inequalities (part B).

\section{Definitions and Main Results}
\begin{figure}[htbp]
  \centering
  \includegraphics[width=0.8\textwidth]{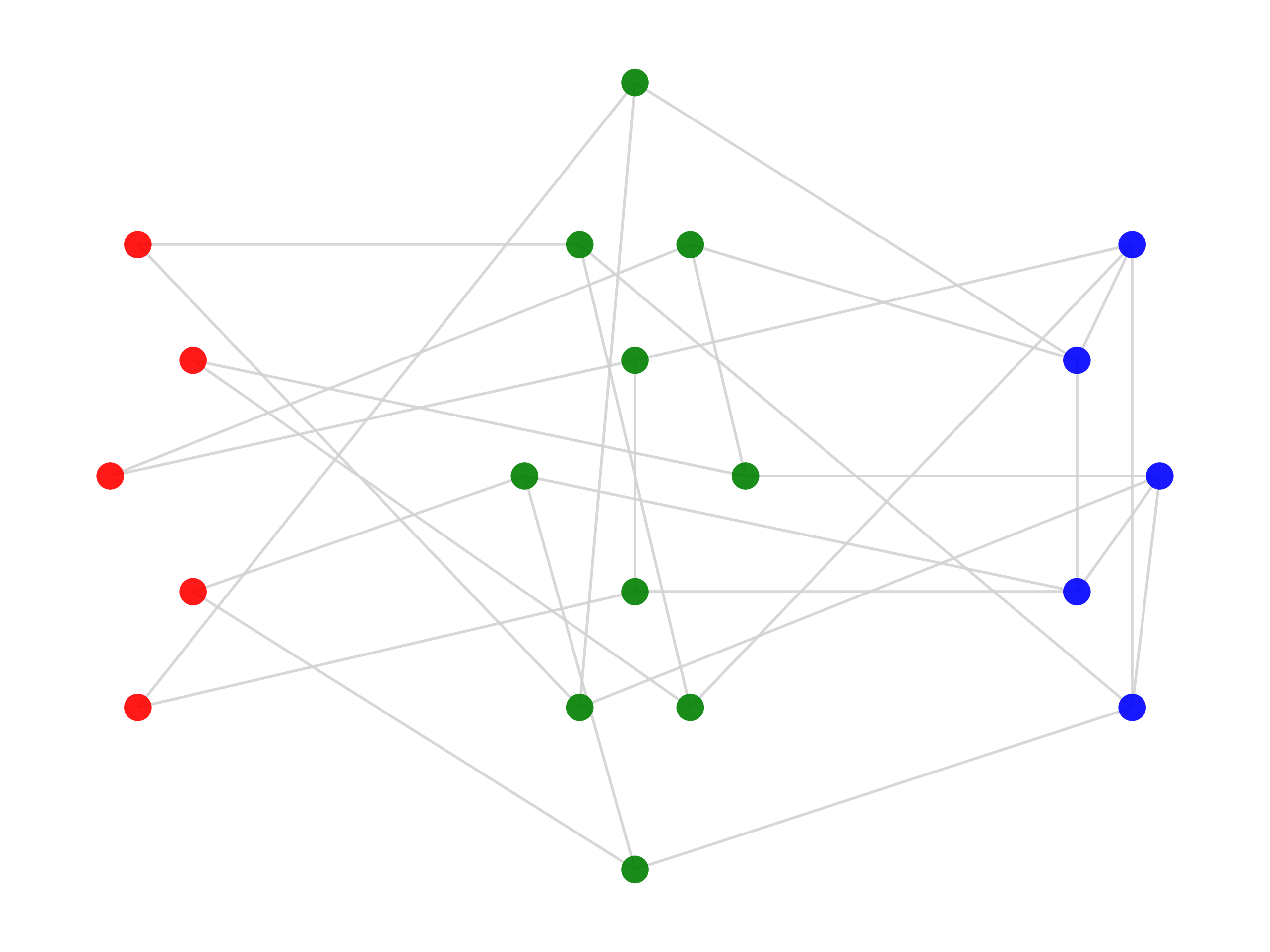} 
  \caption{An example of a $\boldsymbol{k}$-regular graph with 3 classes of respectively red, green and blue nodes; $\boldsymbol{k}=\bigg(\begin{smallmatrix}
   0&  2& 0\\
   1& 1 & 1\\
   0&  2& 2
  \end{smallmatrix}\bigg)$, $\boldsymbol{\alpha}=(0.25,0.5,0.25)$, $N=20$.   }
  \label{fig: kregular graph}
\end{figure}

Let us fix $n\in\mathbb{N}$, an irreducible integer matrix  $\mathbf{k}= (k_{ab}\in \mathbb{N})_{a,b\leq n}$ and a positive vector $\boldsymbol{\alpha}= ( \alpha_a \in \mathbb{Q}^+)_{a\leq n}$ satisfying the following constraints:
    \begin{itemize}
        \item[i)] $\sum_{a=1}^n \alpha_a = 1\,$,  \label{cond1}
        \item[ii)] $\alpha_a\, k_{ab}=\alpha_b\,k_{ba}\ $ for $1\leq a<b\leq n\,$. \label{cond2}
    \end{itemize}
We consider the sequence of integer numbers $N\in\mathbb N$ such that for $a=1,...,n$ 
    \begin{itemize}
        \item[iii)] $N_a:=\alpha_a N$ is integer, 
        \item[iv)] $N_a\,k_{aa}$ is even.
    \end{itemize}
    Let $V_N$ be a set of cardinality $N$. Let $V_N^{(1)},...,V_N^{(n)}$ be a partition of $V_N$ in $n$ non-empty classes of cardinalities $N_1,...,N_n$ respectively.
    It is convenient to assume $V_N^{(a)}\subset V_{N'}^{(a)}$ for $N<N'$, and set $C_a:=\bigcup_{N} V_N^{(a)}$ for $a=1,\dots,n$.

\begin{definition}\label{kregular}     
    A graph on the vertex set $V_N$ is called \textit{$\mathbf{k}$-regular} if for every couple $a,b=1,...,n$, every vertex of each class $C_{a}$ has exactly $k_{ab}$ neighbours of class $C_{b}$.
    In this case $k_a:=\sum_{b=1}^n k_{ab}$ denotes the degree of each vertex of class $C_a$, $a=1,\dots,n$.
\end{definition}

\begin{definition}
    A \textit{random $\mathbf{k}$-regular graph} $G_N=(V_N,E_N)$ is uniformly drawn among the set of all $\mathbf{k}$-regular graphs with the same vertex set and the same partition in classes. The corresponding probability measure will be denoted by $\mathbb P$.
\end{definition}

\begin{remark}
A random $\mathbf{k}$-regular graph on the vertex set $V_N$ can be constructed by the following procedure. For every couple $a,b=1,\dots,n$, for every vertex in $C_a$ one draws exactly $N_a\,k_{ab}$ half-edges, for every vertex in $C_b$ one draws exactly $N_b\,k_{ba}$ half-edges. The former half-edges are successively closed choosing uniformly at random one of the remaining latter half-edges. Notice that conditions ii), iv) ensure that the process is feasible and there are no remaining half-edges. 
\end{remark}

It is convenient to introduce another graph representing the connections among classes: let us denote by $\mathcal G$ the graph with vertex set $\{1,\dots,n\}$ and weighted adjacency matrix $\mathbf{k}$. Edges of $\mathcal G$ are direct, weighted and self-edges may be present. The set of edges of $\mathcal G$ is
\begin{equation}
   \mathcal{E} := \big\{(a,b)\in\{1,\dots,n\}^2:\, k_{ab}\neq 0 \big\} \,=\, \big\{(a,b)\in\{1,\dots,n\}^2:\, k_{ba}\neq 0 \big\} 
\end{equation}
thanks to property \textrm{ii)} above. We denote by $\mathcal{E}_a:=\{b:\, (a,b)\in\mathcal{E}\}$ the set of neighbours of $a=1,\dots,n$. Throughout the paper we will always assume that the matrix $\mathbf k$ is irreducible, so that the graph $\mathcal G$ is strongly connected.

\begin{definition} Let $\boldsymbol{\beta}=(\beta_{ab}\in\mathbb R^+)_{(a,b)\in \mathcal E,\,a\leq b}$ and $\boldsymbol{h}=(h_a\in\mathbb R)_{a\leq n}$, called respectively \textit{coupling coefficients} and \textit{external fields}.
An Ising  model on the random $\mathbf{k}$-regular graph $G_N= (V_N,E_N)$ is described by the following (random) Hamiltonian 
\begin{equation}\label{Hamiltonian}
    H_{G_N}(\boldsymbol{\sigma}) \,:=\, -\sum_{\substack{(a,b)\in \mathcal E:\\a\leq b}} \beta_{ab} \sum_{\substack{ij\in E_N:\\i\in C_a,\,j\in C_b\\ \textrm{or viceversa}}}\sigma_i\sigma_j \,-\, \sum_{a\leq n} h_a\sum_{\substack{i \in V_N:\\i\in C_a}}\sigma_i
\end{equation}
for $\boldsymbol{\sigma}=(\sigma_i)_{i\in V_N}\in\{-1,1\}^{V_N}$. Without loss of generality, we will write $\beta_{ab}\equiv\beta_{ba}$\;for every $(a,b)\in \mathcal E$.
\end{definition}

The Boltzmann-Gibbs expectation induced by the Hamiltonian \eqref{Hamiltonian} will be denoted by $\langle \cdot\rangle_{G_N,\boldsymbol \beta,\boldsymbol h}$ or simply $\langle \cdot\rangle_{G_N}$.
Notice that this is a random measure, due to the randomness of the graph $G_N$. 

The main result of the present paper is the characterization of the large $N$ limit of the main physical quantities at equilibrium: the magnetisation, the internal energy and the free energy per particle. The large $N$ limit is taken along a sequence satisfying conditions iii)  and iv).
As the observables are random quantities, we will prove convergence in $L^r(\mathbb P)$-norm w.r.t. the randomness of the graph $G_N$.

\begin{theorem}[Positive external fields]\label{main:magnetization}
    Consider the Ising model \eqref{Hamiltonian} on a random $\mathbf{k}$-regular graph $G_N$.
    Assume one of the following conditions:
    \begin{itemize}
        \item[i)] $h_1,\dots,h_n>0\,$;
        \item[ii)]  $h_1,\dots,h_n\geq0$ at least one of them striclty positive, and $k_1,\dots, k_n\geq 2\,$.
    \end{itemize}   
    Then for every vertex $i\in C_a$, $a=1,\dots,n$ the (random) spin magnetization converges in $L^r(\mathbb P)$-norm for any $r\in[1,\infty)$ to
    \begin{equation}
    \langle\sigma_i\rangle_{G_N} \quad \xrightarrow[N\rightarrow\infty]{L^r(\mathbb P)} \quad  \tanh\Big( h_a \,+\, \sum_{c\in\mathcal E_a} k_{ac}\, F_{\beta_{ac}}\big(\bar{z}_c^{(a)}\big) \Big) \;,
    \label{eq: main magnetization}
    \end{equation}
    and for every edge $ij\in E_N$, $i\in C_a,\,j\in C_{a'}$ with $(a,a')\in\mathcal E$, the (random) two points correlation converges to
    \begin{equation}
           \langle\sigma_i\sigma_j\rangle_{G_N} \quad \xrightarrow[N\rightarrow \infty]{L^r(\mathbb P)} \quad
           \tanh\Big(  \beta_{aa'} \,+\, \tanh^{-1}\big( \tanh(\bar{z}_a^{(a')})\tanh(\bar{z}_{a'}^{(a)}) \big) \Big) \;,           
    \label{eq: main two point correlation}
    \end{equation}
    where
    \begin{equation}
        F_\beta(x) := \tanh^{-1}(\tanh(\beta)\tanh(x)) \;,
    \label{F_beta definition}
    \end{equation} $\boldsymbol{\bar z} = (\bar{z}_a^{(b)})_{(a,b)\in\mathcal E}$ is the unique fixed point with non-negative (actually positive) components of the following multi-dimensional recursion
    \begin{equation} \label{recursion main}
    z_a^{(b)} (s) \,=\, h_a + \sum_{c\in \mathcal E_a} (k_{ac}-\delta_{bc})\, F_{\beta_{ac}}\big(z_c^{(a)} (s-1)\big)\,, \quad (a,b)\in\mathcal E\,, \quad s\geq1
    \end{equation}
and $\delta_{bc}$ denotes the Kronecker delta.
Moreover, $\boldsymbol{\bar{z}}$ is reached as $s\to\infty$ starting from any non-negative initialization $(z_a^{(b)}(0)\geq0)_{(a,b)\in\mathcal E}$ of the recursion.
\end{theorem}

\begin{remark} \label{remark:convergence_average}
    For every vertex $i\in V_N$ belonging to the same class $C_a$ the spin magnetisation $\langle\sigma_i\rangle_{G_N}$ shares the same distribution under the measure $\mathbb P$. And for every edge $ij\in E_N$ with endpoints belonging to the same two classes $C_a,C_{a'}$ the spin-spin correlation $\langle\sigma_i\sigma_j\rangle_{G_N}$ shares the same distribution.
    Therefore the macroscopic observables corresponding to \eqref{eq: main magnetization}, \eqref{eq: main two point correlation} also converge:
    \begin{equation}
    \frac{1}{N}\sum_{i\in V_N}\langle\sigma_i\rangle_{G_N} \ \xrightarrow[N\rightarrow\infty]{L^r(\mathbb P)} \ \  \sum_{a\leq n}\,\alpha_a\, \tanh\bigg( h_a \,+\, \sum_{c\in\mathcal E_a} k_{ac}\, F_{\beta_{ac}}\big(\bar{z}_c^{(a)}\big) \bigg)
    \label{eq: main magnetization _ average}
    \end{equation}
    \begin{equation}
        \frac{1}{N}\sum_{ij\in E_N}\langle\sigma_i\sigma_j\rangle_{G_N} \ \xrightarrow[N\rightarrow \infty]{L^r(\mathbb P)} \  \sum_{(a,a')\in\mathcal E} \frac{\alpha_a\,k_{aa'}}{2^{\delta_{aa'}}}  \tanh\Big(  \beta_{aa'} \,+\, \tanh^{-1}\big( \tanh(\bar{z}_a^{(a')})\tanh(\bar{z}_{a'}^{(a)}) \big) \Big) 
    \label{eq: main two point correlation _ average}
    \end{equation}
    where the factor $\frac{\alpha_a\,k_{aa'}}{2^{\delta_{aa'}}}$ multiplied by $N$ corresponds to the number of edges connecting a vertex of class $C_a$ with a vertex of class $C_{a'}$.
\end{remark}

\begin{theorem}\label{main:pressure}
    Under the same hypothesis and notation of Theorem \ref{main:magnetization}, consider the (random) partition function
    \begin{equation}  \label{partitionfun}  Z_{G_N}\,:=\sum_{\boldsymbol{\sigma}\in\{-1,1\}^{V_N}} \, e^{-
    H_{G_N}(\boldsymbol{\sigma})} \;.
    \end{equation} 
    Then:
    \begin{equation}
        p_{G_N}:=\,N^{-1}\log Z_{G_N}      \ \xrightarrow[N\rightarrow \infty]{L^r(\mathbb P)} \      p(\boldsymbol{\beta},\boldsymbol{h})\;,
 \label{eq_free_en_convergence}
    \end{equation}
    \begin{equation} \begin{split}
        p(\boldsymbol{\beta},\boldsymbol{h}) :=\,& -\frac{1}{2}\, \sum_{a\leq n} \alpha_a \sum_{a'\in \mathcal E_a} k_{aa'} \bigg( \frac{\log(1-\theta_{aa'}^2)}{2} \,+\, \log\big(1+\theta_{aa'}\,\bar{m}_a^{(a')}\,\bar{m}_{a'}^{(a)}\big)\bigg) \,+ \\ 
        &+\,\sum_{a\leq n} \alpha_a\,\log\bigg(e^{h_a} \prod_{a'\in \mathcal E_a} \big(1+\theta_{aa'}\,\bar{m}_{a'}^{(a)}\big)^{k_{aa'}} +\, e^{-h_a} \prod_{a'\in \mathcal E_a} \big(1-\theta_{aa'}\,\bar{m}_{a'}^{(a)})\big)^{k_{aa'}}\bigg)
        \label{free_energy}
    \end{split} \end{equation}
    where we set $\theta_{aa'}:=\tanh(\beta_{aa'})$ and $\bar m_{a}^{(a')}:=\tanh(\bar z_a^{(a')})$.
\end{theorem}

    Given the matrices $\mathbf{k},\boldsymbol{\beta}$, let us introduce the \textit{weighted non-backtracking} matrix $\boldsymbol M$, indexed by $\mathcal E \times \mathcal E$, with entries
    \begin{equation}
        M_{(a,b),(c,d)} \,:=\, \begin{cases}
            \,\tanh(\beta_{ac}) \,(k_{ac}-1) &\textrm{\,if } a=d, b=c \\
            \, \tanh(\beta_{ac}) \,k_{ac} &\textrm{\,if }a=d, b\neq c \\
            \, 0 &\textrm{\,if }a\neq d
        \end{cases}
    \label{M: definition}
    \end{equation}  
    for every $(a,b)\in \mathcal E$, $(c,d)\in \mathcal E$.
The name is justified by the similarity with the \textit{non-backtracking matrix} (see \cite{kempton2016non,glover2020spectral}). In the following, we indicate by $\rho(\boldsymbol M)$ the spectral radius of $\boldsymbol M$.

\begin{theorem}[High temperature, arbitrary external fields]\label{main: high temperatures}
    Consider the Ising model \eqref{Hamiltonian} on a random $\mathbf{k}$-regular graph $G_N$. 
    If $\rho(\boldsymbol M)<1$, then the spin magnetisations
    verify \eqref{eq: main magnetization}, 
    now denoting by $\boldsymbol{\bar z} = (\bar{z}_a^{(b)})_{(a,b)\in\mathcal E}$ the unique fixed point in $\mathbb R^\mathcal E$ of the recursion \eqref{recursion main} reached as $s\to\infty$ from any initialization.
\end{theorem}

For the next Theorem, we need to exclude the particular case in which the graph of classes $\mathcal{G}$ is made of only two cycles, oppositely directed, and in at least one of them a walk can never backtrack (i.e. $\mathcal{G}$ is \textit{"simply cyclic"}, see Definition \ref{def: non-cyclic } in the Appendix).


\begin{theorem}[Phase transition at zero external fields]\label{main: phase transition}
Consider the Ising model \eqref{Hamiltonian} on a random $\mathbf{k}$-regular graph $G_N$.
Assume $k_1,\ldots,k_n\geq 2$ and $\mathcal G$ not simply cyclic.
For simplicity assume homogeneous external field $h:=h_1=\ldots=h_n$.
The spontaneous magnetization of a spin $\sigma_i$, $i\in C_a$, can be defined as
\begin{equation} \label{eq: spontaneous}
    S_a(\boldsymbol{\beta}) :=\, \lim_{h\to 0^+}\lim_{N\to\infty}\langle \sigma_i\rangle_{G_N,\boldsymbol \beta,h} 
\end{equation}
where the limit with respect to $N$ is in $L^r(\mathbb P)-$norm. We have:
\begin{equation}
    S_a(\boldsymbol{\beta}) \,=\, \begin{cases}
        \,0 &\textrm{,\,if } \rho(\boldsymbol M)<1 \\
        \,\tanh\Big( \sum_{c\in\mathcal E_a} k_{ac}\, F_{\beta_{ac}}\big(\bar{z}_c^{(a)+}\big) \Big)>0 &\textrm{,\,if }\rho(\boldsymbol M)> 1 
    \end{cases}
\label{S zero fields function}
\end{equation}
where $\boldsymbol{\bar z^+} = (\bar{z}_a^{(b)+})_{(a,b)\in\mathcal E}$ is the unique fixed point with all positive components of the recursion \eqref{recursion main} with $h_1=\ldots=h_n=0$.
\end{theorem}

Being a non-analytic function, $S_a(\boldsymbol \beta)$ denotes the presence of a phase transition in the model when $h=0$ and $\rho(\boldsymbol M)=1\,$.

\begin{remark}\label{main: remark phase transition}
In the case of homogeneous $\beta:=\beta_{ab}$ for all $(a,b)\in\mathcal E$, the critical inverse temperature $\beta=\beta_{\textrm c}$ such that $\rho(\boldsymbol M)=1$ rewrites more explicitly as 
\begin{equation}
    \beta_{\textrm c} \,=\, \tanh^{-1}\bigg(\frac{1}{\rho({\boldsymbol {\bar M}})}\bigg)
\end{equation}
where the entries of $\boldsymbol {\bar M}$ are
\begin{equation}
    \bar M_{(a,b),(c,d)} \,:=\, \begin{cases}
            \,k_{ac}-1 &\textrm{\,if } a=d, b=c \\
             \,k_{ac} &\textrm{\,if }a=d, b\neq c \\
            \, 0 &\textrm{\,if }a\neq d
        \end{cases} \;.
    \label{M: definition}
    \end{equation}  
for $(a,b),\,(c,d)\in\mathcal E$.
The following bounds hold true:
    \begin{equation}
        \tanh^{-1}\bigg(\frac{1}{\max_{a=1,\ldots,n} k_a-1}\bigg) \,\leq\, \beta_{\textrm c} \,\leq\, \tanh^{-1}\bigg(\frac{1}{\min_{a=1,\ldots,n} k_a-1}\bigg) \;.
    \label{eq bounds critical temperature}
    \end{equation}
In particular for $k$-regular random graphs, i.e., for $n=1$, we find $\beta_{\textrm c} = \tanh^{-1}(\frac{1}{k-1})$ in agreement with \cite{dembo2010ising,leone2002ferromagnetic}. 
In \eqref{eq bounds critical temperature}, the lower bound is obtained by taking the $\infty$-norm of $\boldsymbol {\bar M}$, while for the upper bound one follows the proof of Corollary \ref{prop: zero fields rho(M)>1}, applying a small homogeneous boundary field.
\end{remark}

\section{General results on trees} \label{sec:trees}

\subsection{Recursion}

Let us consider a finite \textit{tree} $T=(V(T),E(T))$ and a general Ising model on $T$ with Hamiltonian 
\begin{equation}
    H_T(\boldsymbol{\sigma}) \,=\, -\sum_{ij\in E(T)}\beta_{ij}\,\sigma_i\sigma_j - \sum_{i\in V(T)}h_i\,\sigma_i \;.
\label{general hamiltonian}
\end{equation}
Denote by $\mu_T(\cdot)$, $\langle\cdot\rangle_T$ the associated Gibbs measure and expectation.
In this subsection we recall a well-known recursion on trees, then we apply it to the $\mathbf{k}$-regular case.

Fix a root $v$ among the vertices of the tree $T$.
Given a vertex $w$, we denote by $T_w$ the subtree rooted at $w$ generated by $w$ itself together with the connected vertices that lay further than $w$ from $v$.
The vertices $u_1,\dots,u_{K_w}$ that are neighbours of $w$ in the subtree $T_w$ are called the \textit{children} of $w$. $K_w$ is called the \textit{forward degree} of $w$ and corresponds to $\deg_T(w)-1$, unless $w=v$.

\begin{lemma}\label{lemma_general_tree_recirsion}
    Let $w$ be any vertex of the rooted tree $T$, let ${m}_w := \langle \sigma_w\rangle_{T_w}$ denote its magnetization in the Ising model \eqref{general hamiltonian} restricted to the tree $T_w$.  Set $z_w := \tanh^{-1}({m}_w)$.
    The following recurrence relation holds true:
    \begin{equation}
        z_w \,=\, h_w \,+\, \sum_{i=1}^{K_w} \,F_{\beta_{wu_i}}(z_{u_i})
    \label{lemma1}
    \end{equation}   
    where $u_1,\dots,u_{K_w}$ are the children of $w$ and $F_\beta$ is defined by \eqref{F_beta definition}.
\label{recursion_generic_tree}
\end{lemma}

\begin{remark}\label{remark:root_magnetization}
    Taking $w=v$, then $T_v= T$ and the recursion leads to the root magnetization for the Ising model on the original tree $T$, $m_v = \langle\sigma_v\rangle_T$. 
\end{remark}

\begin{remark}
    If the vertex $u$ belongs to the boundary of the rooted tree $T$, then $T_u=\{u\}$ and thus $z_u = h_u$.
\end{remark}

\begin{lemma} \label{lemma general two points correlation}
    Let $ww'$ be any edge of the tree $T$. Denote by $T_{w}^{(w')},\,T_{w'}^{(w)}$ the two disjoint subtrees rooted respectively in $w,w'$ obtained from $T$ by deleting the edge $ww'$. We have:
    \begin{equation}
        \langle\sigma_w\sigma_{w'}\rangle_T  =  \tanh\Big(  \beta_{ww'} + \tanh^{-1}( m_w^{(w')}\,m_{w'}^{(w)} ) \Big)
    \label{eq general two points}
    \end{equation}
    where $m_w^{(w')}:=\langle  \sigma_w\rangle_{T_{w}^{(w')}}$ is the magnetization of the vertex $w$ in the Ising model \eqref{general hamiltonian} restricted to the tree $T_{w}^{(w')}$.
\end{lemma}

For completeness, the proofs of Lemma \ref{lemma_general_tree_recirsion} and Lemma \ref{lemma general two points correlation} can be found in Appendix \ref{appendix: auxiliary}.

Now we introduce $\mathbf k$-regular trees and specify the recursion \eqref{lemma1} on this class of trees.

\begin{definition} \label{kregulartree}
    Let $a=1,\dots,n$, let $t\in\mathbb N$. A \textit{$\mathbf{k}$-regular tree} of depth $t$ rooted at a vertex of class $C_a$ is denoted by $T_a(\mathbf{k},t)$ (or $T_a(t)$ for simplicity) and defined as follows.
    The root originates $k_{ab}$ children of class $C_b$ for every $b=1,...,n$ ($1$\textsuperscript{st} generation).
    Recursively, for every $b,c=1,\dots,n$ each vertex of class $C_b$ in the $g$\textsuperscript{th} generation whose parent belongs to class $C_{c}$ originates $k_{bc}-1$ children of class $C_c$ and $k_{bd}$ children of class $C_d$ for $d=1,...,c-1,c+1,..,n$ ($(g+1)$\textsuperscript{th} generation).
    This procedure stops at the $t$\textsuperscript{th} generation.  
\end{definition}

\begin{remark} \label{iso1} A finite $\mathbf{k}$-regular tree is not a $\mathbf{k}$-regular graph, because the vertices of the last generation (i.e., the boundary) do not have the prescribed degrees.
Unlikely random $\mathbf{k}$-regular graphs, the realization of a $\mathbf{k}$-regular tree $T_a(t)$ is unique up to graph isomorphism, given the root class and the depth.
\end{remark}

\begin{definition}\label{kregularsubtree}
Let $(a,b)\in\mathcal E$, $s\in\mathbb N$. We denote by $T_{a}^{(b)}(s)$ a tree of depth $s$ rooted at a vertex of class $C_a$, which differs from the $\mathbf{k}$-regular tree $T_a(s)$ only for the fact that the root has $k_{ab}-1$ children of class $C_b$.
\end{definition} 

\begin{remark} \label{iso2}
For each pair $(a,b)$ in $\mathcal E$, we say that a vertex of the tree $T_a(t)$ is of type $(b(c), s)$, meaning that $b$ is the class of the vertex, $c$ is the class of its parent, and $s=0,\dots,t-1$ its distance from the boundary of the tree.
Every subtree of $T_a(t)$ generated by a vertex of type $(b(c), s)$ together with its descendants is isomorphic $T_{b}^{(c)}(s)$.
\end{remark}
\begin{figure}[htbp]
  \centering
  \includegraphics[width=0.7\textwidth]{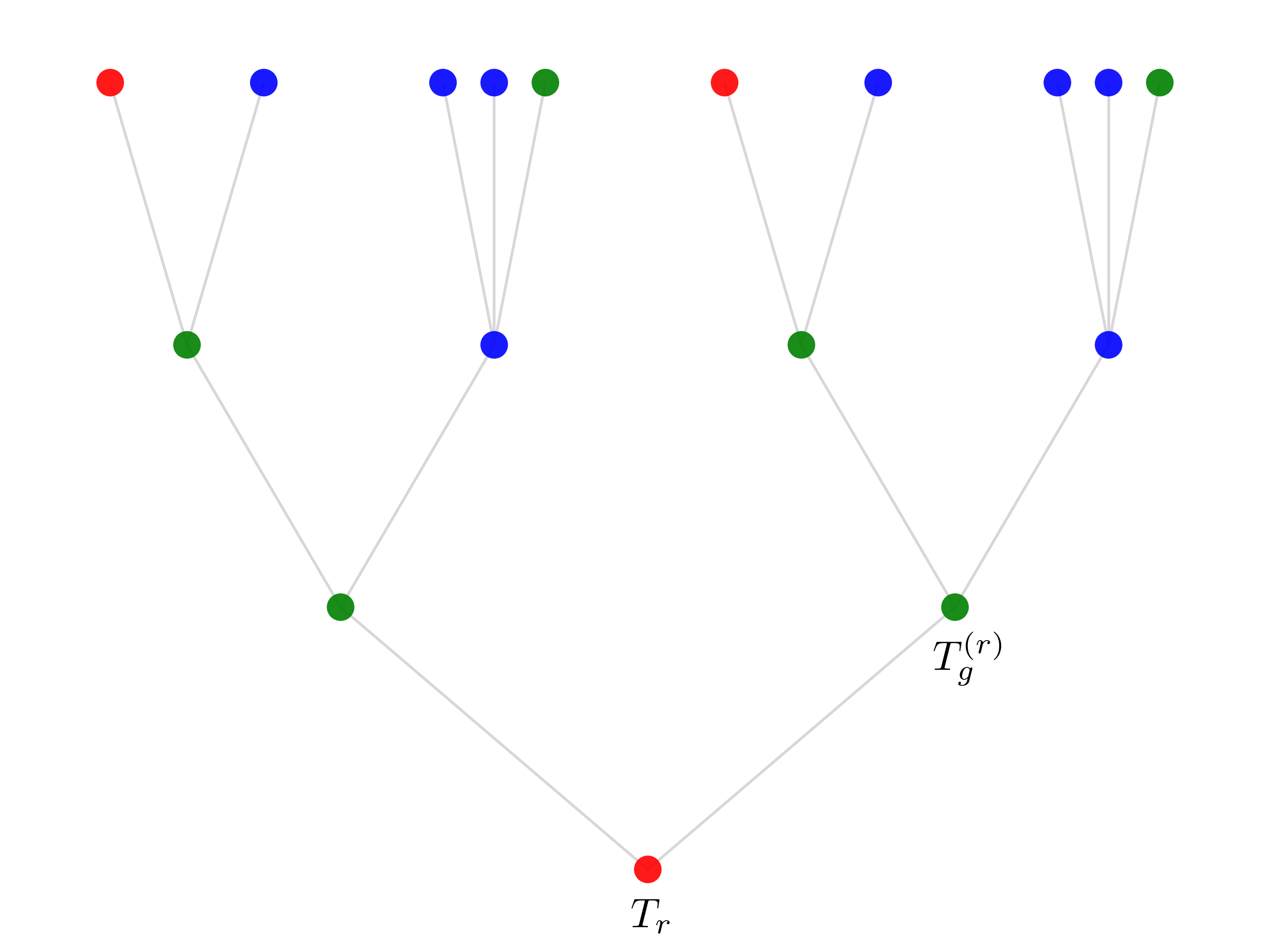} 
  \caption{ One of the $\boldsymbol{k}$-regular trees associated to the class of random $\boldsymbol{k}$-regular graphs in Fig. \ref{fig: kregular graph}, namely $T_r(3)$, is shown. $T_g^{(r)}(2)$ is also indicated.}
  \label{fig: kregular tree}
\end{figure}

For the rest of Section \ref{sec:trees} we will consider an Ising model on the $\mathbf{k}$-regular tree $T\equiv T_a(t)$ or $T\equiv T_{a}^{(b)}(t)$
with class-dependent couplings and fields, up to an extra field $\boldsymbol{h_*}=(h^{(b)}_{*a}\in[-\infty,\infty])_{(a,b)\in\mathcal E}$ addedd at the boundary $T_*$, that is the set of vertices at distance $t$ from the root. Precisely we consider the following  Hamiltonian:
\begin{equation}\label{perturbed hamiltonian}
    H_{T}(\boldsymbol{\sigma}) \,= -\sum_{\substack{(a,b)\in \mathcal E:\\a\leq b}} \beta_{ab} \sum_{\substack{ij\in E(T):\\i\in C_a,j\in C_b\\\textrm{or viceversa}}}\sigma_i\sigma_j \,-\, \sum_{a\leq n}\,h_a\sum_{\substack{i\in V(T):\\i \in C_a}}\sigma_i \,- \sum_{(a,b)\in\mathcal E}h^{(b)}_{*a}\sum_{\substack{i\in  T_*:\\i \in C_a\,\partial i \in C_b}}\sigma_i \;.
\end{equation}
where $\partial i$ identifies the parent of vertex $i$ in the tree $T$.
Notice that the last generations of $T$ can be empty, in that case $T_*$ is also empty.

\begin{proposition}
    Let ${m}^{(b)}_a(s,\boldsymbol{h_*})$
    denote the magnetization of the root in the Ising model \eqref{perturbed hamiltonian} on the tree $T\equiv T_{a}^{(b)}(s)$.
    Set $z^{(b)}_a(s,\boldsymbol{h_*}) := \tanh^{-1}({m}^{(b)}_a(s,\boldsymbol{h_*}))$.
    The following recursion holds true for every $(a,b)\in\mathcal E$
    \begin{equation} \left\{
    \begin{split}
        &\,z_a^{(b)} (s,\boldsymbol{h_*}) \,=\, h_a \,+\, \sum_{c\in\mathcal E_a}\, (k_{ac}-\delta_{bc})\,F_{\beta_{ac}}\big(z_c^{(a)} (s-1,\boldsymbol{h_*}) \big) \,,\quad s\geq1 \\
        &\,z_a^{(b)} (0,\boldsymbol{h_*}) \,=\, h_a + h_{*a}^{(b)} \,,\quad s=0
    \end{split} \right.
    \label{recursion_regular_trees}
    \end{equation}   
    where $F_\beta(x)$ is defined in (\ref{F_beta definition}) and $\delta_{bc}$ is the Kronecker delta.
    
    Moreover, let $m_a(t,\boldsymbol{h_*})$
    denote the magnetization of the root in the Ising model \eqref{perturbed hamiltonian} on the $\mathbf{k}$-regular tree $T\equiv T_a(t)$. Set $z_a(t,\boldsymbol{h_*}):= \tanh^{-1}(m_a(t,\boldsymbol{h_*}))$. We have:
    \begin{equation}
    \label{recursion_regular_trees_root}
    z_a(t,\boldsymbol{h_*}) \,=\, h_a \,+\, \sum_{c\in\mathcal E_a}\,k_{ac}\,F_{\beta_{ac}}\big(z_c^{(a)} (t-1,\boldsymbol{h_*})\big) \;.
    \end{equation}
\end{proposition} 

\begin{proof}
    By Remarks \ref{iso1}, \ref{iso2}, the proof is a streightforward application of Lemma \ref{recursion_generic_tree} to the Ising model \eqref{perturbed hamiltonian} on a $\mathbf{k}$-regular tree $T_a(t)$, as long as $s<t$. 
\end{proof}

In the following we will denote by $\boldsymbol{z}(s,\boldsymbol{h_*}) = \big(z_a^{(b)}(s,\boldsymbol{h_*})\big)_{(a,b)\in\mathcal E}$ the solution of recursion \eqref{recursion_regular_trees} at step $s$.

\subsection{Existence of fixed points}
We show two simple properties of the recursion \eqref{recursion_regular_trees}, then we focus on a condition which provides the existence of fixed points by monotonicity.

\begin{lemma}\label{corollary_bounds_H}
The following bounds hold true
    \begin{equation}
        h_a-\sum_{c\in\mathcal E_a} (k_{ac}-\delta_{bc})\, \beta_{ac} \,\leq\, z_a^{(b)}(s,\boldsymbol{h_*}) \,\leq\,  h_a + \sum_{c\in\mathcal E_a} (k_{ac}-\delta_{bc})\, \beta_{ac} 
    \label{bounds}    
    \end{equation}
for every $(a,b)\in\mathcal E$, $s\geq1$. 
\end{lemma}

\begin{proof}
    It follows from expression \eqref{recursion_regular_trees} using the fact that $|F_{\beta}(x)|\leq \beta$.
\end{proof}

\begin{lemma}
    The function $h_{*d}^{(e)} \mapsto z_a^{(b)}(s,\boldsymbol{h_*})$ is monotonic non-decreasing for every $(a,b),(d,e)\in\mathcal E$, $s\in\mathbb N$. 
\label{corollary_monotony_H}
\end{lemma}

\begin{proof}
    Let us reason by induction on $s$. At the boundary $s=0$ we have
   \begin{equation}
    \frac{\partial z^{(b)}_a(0,\boldsymbol{h_*})}{\partial h_{*d}^{(e)}} \,=\,\delta_{(a,b),(d,e)}\geq0 \;.
   \end{equation}
   At $s\geq1$ suppose that $\frac{\partial z^{(b)}_a(s-1,\boldsymbol{h_*})}{\partial h_{*d}^{(e)}} \geq 0$ for all $(a,b)\in\mathcal E$, then differentiating the recursion \eqref{recursion_regular_trees} we find
    \begin{equation}
    \frac{\partial z^{(b)}_a(s,\boldsymbol{h_*})}{\partial h_{*d}^{(e)}} \,=\, \sum_{c\in\mathcal E_a}(k_{ac}-\delta_{bc})\,F_{\beta_{ac}}'\big(z_c^{(a)}(s-1,\boldsymbol{h_*})\big)\,\frac{\partial z^{(a)}_c(s-1,\boldsymbol{h_*})}{\partial h_{*d}^{(e)}}\,\geq0 
   \end{equation}
    since $F'_\beta(x)\geq0$.
\end{proof}

\begin{remark}
   The monotonicity with respect to the extra boundary fields also follows by general FKG inequalities (see Remark \ref{FKG_field} in the Appendix). Indeed $\frac{\partial m_a^{(b)}(s,\boldsymbol{h_*})}{\partial h_{*d}^{(e)}} \geq0$ independently of the sign of the external fields, since the couplings are non-negative.
\end{remark}

Now, provided that the following limits exist, we denote
\begin{equation}
    \bar{z}_a^{(b)}(\boldsymbol{h_*}) := \lim_{s\rightarrow \infty}z_a^{(b)}(s,\boldsymbol{h_*}) \quad \quad \forall \,(a,b)\in\mathcal E
\label{notation_fixed_point}
\end{equation}
the components of the fixed point $\boldsymbol{\bar{z}}(\boldsymbol{h_*})$ of the recursion (\ref{recursion_regular_trees}) reached starting from extra boundary field $\boldsymbol{h_*}$. We simply write $\boldsymbol{\bar{z}}(h_*)$ when the extra boundary field is homogeneous, that is $h_{*a}^{(b)}=h_*$ for every $(a,b)\in\mathcal E$.

The following Proposition uses monotonicity to prove the existence of fixed points for specific choices of extra boundary fields.

\begin{proposition}[Existence of fixed points]
    Let $\boldsymbol{\underline h}=(\underline h_a^{(b)}\geq-\infty)_{(a,b)\in\mathcal E}$ such that
    \begin{equation}
         \sum_{c\in\mathcal E_a}\,(k_{ac}-\delta_{bc})\,F_{\beta_{ac}}(h_c+\underline{h}_{c}^{(a)}) \,\geq\, \underline h_{a}^{(b)} \quad \forall \,(a,b)\in\mathcal E\,.
    \label{condition_H_small}
    \end{equation}
    Notice that $\boldsymbol{\underline h}=-\boldsymbol{\infty}$ is always a feasible choice. 
    Then the sequence $s\mapsto z_a^{(b)}(s,\boldsymbol{\underline h})$ is monotonic non-decreasing for every $(a,b)\in\mathcal E$.
    In particular, a fixed point $\boldsymbol{\bar{z}}(\boldsymbol{\underline h})$ of the recursion \eqref{recursion_regular_trees}  exists and its components satisfy $\bar{z}_a^{(b)}(\boldsymbol{\underline h}) \,\geq\, h_a + \underline h_{a}^{(b)}$.
    
   Similarly, let $\boldsymbol{\overline h}=(\overline h_a^{(b)}\leq\infty)_{(a,b)\in\mathcal E}$ such that
    \begin{equation}
        \sum_{c\in\mathcal E_a}\,(k_{ac}-\delta_{bc})\,F_{\beta_{ac}}(h_c+\overline h_{c}^{(a)}) \,\leq\, \overline h_{a}^{(b)}  \quad \forall \,(a,b)\in\mathcal E\;.
    \label{condition_H_big}
    \end{equation}
    Notice that $\boldsymbol{\overline h}=+\boldsymbol{\infty}$ is always a feasible choice.
    Then the sequence $s\mapsto z_a^{(b)}(s,\boldsymbol{\overline h})$ is monotonic non-increasing for every $(a,b)\in\mathcal E$.
    In particular, a fixed point $\boldsymbol{\bar{z}}(\boldsymbol{\overline h})$ of the recursion \eqref{recursion_regular_trees} exists and its components satisfy
    $\bar{z}_a^{(b)}(\boldsymbol{\overline h}) \,\leq\, h_a + \overline h_{a}^{(b)}$.

    Finally, if $\boldsymbol{\underline h}\,\preceq\,\boldsymbol{h_*}\,\preceq\,\boldsymbol{\overline h}$ where $\preceq$ denotes componentwise inequality, then:
    \begin{equation}
    \boldsymbol{\bar{z}}(\boldsymbol{\underline h}) \,\preceq\, \liminf_{s\rightarrow \infty}\boldsymbol{z}(s,\boldsymbol{h_*}) \,\preceq\, \limsup_{s\rightarrow \infty}\boldsymbol{z}(s,\boldsymbol{h_*}) \,\preceq\, \boldsymbol{\bar{z}}(\boldsymbol{\overline h}) \,.
    \label{squeeze}
    \end{equation}
\label{lemma existence fixed point H_small H_big}
\end{proposition}

\begin{proof}
Condition \eqref{condition_H_small} simply rewrites as
\begin{equation} \label{mono1}
    \boldsymbol{z}(1,\boldsymbol{\underline h})\,\succeq\, \boldsymbol{z}(0,\boldsymbol{\underline h}) \;.
\end{equation}
Therefore we claim that
\begin{equation} \label{monos}
    \boldsymbol{z}(s,\boldsymbol{\underline h}) \,\succeq\, \boldsymbol{z}(s-1,\boldsymbol{\underline h}) \quad \forall \,s\geq1 \,.
\end{equation}
Let us prove the claim by induction on $s$. 
If $\boldsymbol{z}(s-1,\boldsymbol{\underline h})\succeq \boldsymbol{z}(s-2,\boldsymbol{\underline h})$, then applying the recursion \eqref{recursion_regular_trees}, since $F_\beta(x)$ is non-decreasing, one finds:
\begin{equation} \begin{split}
    z_a^{(b)}(s,\boldsymbol{\underline h}) \,&=\, h_a + \sum_{c\in\mathcal E_a}\, (k_{ac}-\delta_{bc})\,F_{\beta_{ac}}\big(z_c^{(a)} (s-1,\boldsymbol{\underline h}) \big) \,\geq\\
    &\geq\, h_a + \sum_{c\in\mathcal E_a}\, (k_{ac}-\delta_{bc})\,F_{\beta_{ac}}\big(z_c^{(a)} (s-2,\boldsymbol{\underline h}) \big) \,=\, z_a^{(b)}(s-1,\boldsymbol{\underline h})
\end{split} \end{equation}
for every $(a,b)\in\mathcal E$. This proves the non-decreasing monotonicity \eqref{monos}.
As a consequence $\boldsymbol{\bar z}(\boldsymbol{\underline h})=\lim_{s\to\infty}\boldsymbol{\bar z}(s,\boldsymbol{\underline h})$ exists, is bounded below by $\boldsymbol{z}(0,\boldsymbol{\underline h})$ and bounded above by \eqref{bounds}.
    
In an analogous way, we can show that condition \eqref{condition_H_big} implies the existence of $\boldsymbol{\bar z}(\boldsymbol{\overline{h}})=\lim_{s\to\infty}\boldsymbol{\bar z}(s,\boldsymbol{\overline{h}})$ by non-increasing monotonicity, bounded above by $\boldsymbol{ z}(0,\boldsymbol{\overline{h}})$ and bounded below by \eqref{bounds}.

Finally, thanks to the monotonicity with respect to the extra boundary fields (Lemma \ref{corollary_monotony_H}), if $\boldsymbol{\underline{h}}\,\preceq\,\boldsymbol{h_*}\,\preceq\,\boldsymbol{\overline{h}}$ then
    \begin{equation}
    \boldsymbol z(s,\boldsymbol{\underline{h}}) \,\preceq\, \boldsymbol z(s,\boldsymbol{h_*}) \,\preceq\, \boldsymbol z(s,\boldsymbol{\overline{h}})
    \end{equation}
for every $s\in\mathbb N$. As $s\to\infty$ we obtain the bounds \eqref{squeeze}.
\end{proof}

\begin{corollary} \label{corollary_existence}
    Suppose $h_1,\dots,h_n\geq0$.
    The homogenous extra boundary fields $\underline h=0$ (free boundary conditions) and $\overline h=+\infty$ (positive boundary conditions) are such that the magnetizations are respectively monotonic non-decreasing and monotonic non-increasing with respect to the inclusion of a new generation.
\end{corollary}

\begin{proof} It follows by Proposition \ref{lemma existence fixed point H_small H_big}. Indeed, if the external fields are non-negative the condition (\ref{condition_H_small}) is satisfied by $\underline h_a^{(b)}=\underline h=0$ for every $(a,b)\in\mathcal E$. On the other hand, condition (\ref{condition_H_big}) is always fulfilled by $\overline h_a^{(b)}=\overline h=+\infty$ for every $(a,b)\in\mathcal E$.
\end{proof}

\begin{remark}
    Corollary \ref{corollary_existence} also follows by general GKS inequalities (see Remark \ref{GKS_rem} in the Appendix). 
\label{remark_proof_GKS}
\end{remark}

\subsection{Uniqueness of fixed points}

In order to fully exploit Proposition \ref{lemma existence fixed point H_small H_big}, we show that $\boldsymbol{\bar{z}}(\boldsymbol{\underline h})= \boldsymbol{\bar{z}}(\boldsymbol{\overline h})$ for suitable choices of $\boldsymbol{\underline h}\,\prec\, \boldsymbol{\overline h}$.

\begin{lemma}
    Suppose there exists $\boldsymbol{\underline h}=(\underline h_a^{(b)})_{(a,b)\in\mathcal E}$ satisfying condition \eqref{condition_H_small} and
    \begin{equation}
        \underline h_a^{(b)} \,\geq\, -h_a \quad\forall\,(a,b)\in\mathcal E \,.
    \label{positive_boundary}
    \end{equation}
    Let $\boldsymbol{h_*}\succeq\boldsymbol{\underline h}$. Then $z_a^{(b)}(s,\boldsymbol h_*)\,\geq\,0$ and
    \begin{equation}
        \frac{\partial^2}{\partial \big(h_{*d}^{(e)}\big)^2}\;z_a^{(b)}(s,\boldsymbol{ h_*}) \,\leq\, 0 
    \label{concavity_condition}
    \end{equation}
    for every $(a,b),\,(d,e)\in\mathcal E$, $s\in\mathbb N$.
\label{lemma_concavity}
\end{lemma}

\begin{proof}
    Proposition \ref{lemma existence fixed point H_small H_big} together with \eqref{positive_boundary} imply that $z_a^{(b)}(s,\boldsymbol{\underline h}) \geq z_a^{(b)}(0,\boldsymbol{\underline h})\geq0$.
    Therefore, by Lemma \ref{corollary_monotony_H} also
    \begin{equation} \label{z_positive}
        z_a^{(b)}(s,\boldsymbol{ h_*})\geq0 \,.
    \end{equation}
    
    Now, let us take the second derivative of recursion (\ref{recursion_regular_trees}) with respect to the extra boundary field $h_{*d}^{(e)}$. For $s\geq1$
    \begin{equation} \label{rec_second_derivative} \begin{split}
        &\frac{\partial^2}{\partial\big(h_{*d}^{(e)}\big)^2}\,z_a^{(b)}(s,\boldsymbol{ h_*}) \,=\, \sum_{c\in\mathcal E_a}\, (k_{ac}-\delta_{bc})\; \frac{\partial^2}{\partial\big(h_{*d}^{(e)}\big)^2}\,F_{\beta_{ac}}\big(z_c^{(a)}(s-1,\boldsymbol{ h_*})\big) \,=\\
        &=\, \sum_{c\in\mathcal E_a}\, (k_{ac}-\delta_{bc})\, \bigg(F_{\beta_{ac}}''   \big(z_c^{(a)}(s-1,h_*)\big)\,\bigg(\frac{\partial}{\partial h_{*d}^{(e)}}\, z_c^{(a)}(s-1,\boldsymbol{ h_*})\bigg)^2 +\\
        &\quad +  F_{\beta_{ac}}'\big(z_c^{(a)}(s-1,\boldsymbol{h_*})\big) \;\frac{\partial^2}{\partial\big(h_{*d}^{(e)}\big)^2}\,z_c^{(a)}(s-1,\boldsymbol{ h_*}) \bigg)
    \end{split} \end{equation}
    and at the boundary $s=0$ we have $\frac{\partial^2}{\partial\big(h_{*d}^{(e)}\big)^2}\, z_a^{(b)}(0,\boldsymbol{h_*}) 
    =0$.
    Observe that $F'_\beta(x)\geq0$ for every $x\in\mathbb R$, while $F''_\beta(x)\leq0$ only for $x\geq0$. Therefore identity \eqref{rec_second_derivative} together with \eqref{z_positive} can be used to prove \eqref{concavity_condition} by induction on $s$.
\end{proof}

\begin{remark}
    For $h_1,\dots,h_n\geq0$, the concavity condition (\ref{concavity_condition}) can be alternatively proved by correlation inequalities, specifically the GHS inequality (see Appendix \ref{remark_proof_GKS}).
\label{remark_concavity_GHS}
\end{remark}

\begin{proposition}[Existence and uniqueness of fixed point]
    Suppose there exists $\boldsymbol{\underline h}=(\underline h_a^{(b)})_{(a,b)\in\mathcal E}$ verifying \eqref{positive_boundary}, \eqref{condition_H_small} and
    \begin{equation}
       \sum_{c\in\mathcal E_a} (k_{ac}-\delta_{bc})\, F_{\beta_{ac}}(h_c+\underline h_c^{(a)}) \,>\, \underline h_a^{(b)} \quad\forall\,(a,b)\in\mathcal E \textrm{ s.t. } k_a\geq2 \;.
    \label{strict_condition_hsmall}
    \end{equation}
    Then for every $\boldsymbol{ h_*} \,\succeq\,\boldsymbol{\underline h}$ we have
    \begin{equation}
        \boldsymbol{\bar{z}}(\boldsymbol{\underline h}) \,=\, \boldsymbol{\bar z}(\boldsymbol{h_*}) \,=\, \boldsymbol{\bar{z}}(+\infty) \,=: \boldsymbol{\bar{z}} \,.
    \label{uniqueness_equation}
    \end{equation}
    $\boldsymbol{\bar{z}}$ is the unique fixed point of recursion \eqref{recursion_regular_trees} having components $\bar{z}_a^{(b)}\geq h_a+\underline h_a^{(b)}$ for every $(a,b)\in\mathcal E$.
    
\label{lemma_uniqueness}
\end{proposition}

\begin{proof}
    For $(a,b)\in\mathcal E$ such that $\sum_{c\in\mathcal E_a}\, (k_{ac}-\delta_{bc})=k_a-1=0$ the uniqueness is trivial since by \eqref{recursion_regular_trees} we have $\bar{z}_a^{(b)}(\boldsymbol{h_*})=z_a^{(b)}(s,\boldsymbol{h_*})=h_a$ for every $s\geq1$, $\boldsymbol{h_*}\in[-\infty,\infty]^{\mathcal E}$.
    From now on we consider $(a,b)\in\mathcal E$ such that $\sum_{c\in\mathcal E_a}\, (k_{ac}-\delta_{bc})=k_a-1>0$.
    
    The functions $\rho \mapsto z_a^{(b)}(s,\,\boldsymbol{\underline h}+\rho\,\boldsymbol{1})$ are concave for $\rho\geq0$ (Lemma \ref{lemma_concavity}) for every $s\in\mathbb N$.
    Thus, setting $\boldsymbol{\overline h}:=\boldsymbol{\underline h} + R\,\boldsymbol{1}$ and $\boldsymbol{\widetilde h}:=\boldsymbol{\underline h} + \rho\,\boldsymbol{1}
    $ for $\rho\in(0,R)$, $R\in(0,\infty)$, we have
    \begin{equation}
        z_a^{(b)}(s,\boldsymbol{\overline h}) - z_a^{(b)}(s,\boldsymbol{\underline h}) \,\leq\, \frac{R}{\rho}\,\big(z_a^{(b)}(s,\boldsymbol{\widetilde h})-z_a^{(b)}(s,\boldsymbol{\underline h})\big)
\label{consequence_concavity}
    \end{equation}
    for every $s\in\mathbb N$.
    Let us choose $R\geq\max_{a,b}\big(\sum_{c\in\mathcal E_a}(k_{ac}-\delta_{bc})\,\beta_{ac} -\underline h_a^{(b)}\big)$, so that $\boldsymbol {\overline h} $ satisfies condition \eqref{condition_H_big} by Lemma \ref{corollary_bounds_H}.
    Let us choose $\rho\in (0,R)$ sufficiently small so that 
    \begin{equation}\label{eq:monos} 
        z_a^{(b)}(s,\boldsymbol {\widetilde h}) \,\leq\, z_a^{(b)}(s+1,\boldsymbol {\underline h})   \quad\forall\,s\in\mathbb N\,.
    \end{equation}
    One can prove that such $\rho$ exists by induction on $s$. Indeed by hypothesis \eqref{strict_condition_hsmall}, for $\rho>0$ sufficiently small we have
    \begin{equation}\label{eq:mono3} \sum_{c\in\mathcal E_a} (k_{ac}-\delta_{bc})\, F_{\beta_{ac}}(h_c+\underline h_c^{(a)})\,\geq\, \underline h_a^{(b)} + \rho \;,
    \end{equation}
    that is
    \begin{equation}\label{eq:monos2}
      z_a^{(b)}(1,\boldsymbol{\underline h}) \,\geq\, z_a^{(b)}(0,\boldsymbol{\widetilde h}) \,.
    \end{equation}
    Then \eqref{monos} follows using  the recursion \eqref{recursion_regular_trees} and the monotonicity of $F_{\beta}$.
    
    Now, the l.h.s. of \eqref{consequence_concavity} can be bounded below uniformly in $s$ using Proposition \ref{lemma existence fixed point H_small H_big}:
    \begin{equation} \label{bound1}
        z_a^{(b)}(s,\boldsymbol{\overline{h}}) - z_a^{(b)}(s,\boldsymbol{\underline{h}}) \,\geq\, z_a^{(b)}(t,\boldsymbol{\overline{h}}) - z_a^{(b)}(t,\boldsymbol{\underline{h}}) \quad\forall\,s\leq t\,.
    \end{equation}
    Plugging inequalities \eqref{bound1} and \eqref{eq:monos} into \eqref{consequence_concavity} we find:
    \begin{equation}
        z_a^{(b)}(t,\boldsymbol{\overline{h}}) - \bar z_a^{(b)}(t,\boldsymbol{\underline{h}}) \,\leq\, \frac{R}{\rho}\, \big( z_a^{(b)}(s+1,\boldsymbol{\underline{h}})-z_a^{(b)}(s,\boldsymbol{\underline{h}})\big) \;,
\label{consequence_concavity_after H tilde}
    \end{equation}
    then summing over $s=0,\dots,t$ we get:
    \begin{equation}
        (t+1)\,\big( z_a^{(b)}(t,\boldsymbol{\overline{h}}) - z_a^{(b)}(t,\boldsymbol{\underline{h}})\big) \,\leq\, \frac{R}{\rho}\,\big(z_a^{(b)}(t+1,\boldsymbol{\underline{h}})-z_a^{(b)}(0,\boldsymbol{\underline{h}})\big)  \,.
\label{sum_consequence_concavity_after H tilde}
    \end{equation}
    The r.h.s. of \eqref{sum_consequence_concavity_after H tilde} is bounded by a suitable constant $C>0$ independent of $t$ (Lemma \ref{corollary_bounds_H}).
    Therefore we have: 
    \begin{equation}
    z_a^{(b)}(t,\boldsymbol{\overline{h}}) - z_a^{(b)}(t,\boldsymbol{\underline{h}}) \,\leq\, \frac{C}{t+1} \,.
    \end{equation}
    The l.h.s. is always non-negative (Lemma \ref{corollary_monotony_H}). Hence letting $t\rightarrow\infty$, we get
    \begin{equation}
    \bar{z}_a^{(b)}(\boldsymbol{\overline{h}}) - \bar{z}_a^{(b)}(\boldsymbol{\underline{h}}) \,=\, 0 \,.
    \end{equation}
    
    Now, using the bounds in \eqref{squeeze}
    it follows that
    \begin{equation} \label{uniqueness_november}
    \boldsymbol{\bar{z}}(\boldsymbol{\underline{h}}) = \boldsymbol{\bar{z}}(\boldsymbol{h_*}) = \boldsymbol{\bar{z}}(\boldsymbol{\overline{h}}) \,=:\, \boldsymbol{\bar{z}}
    \end{equation}
    for any $\boldsymbol{\underline{h}}\preceq \boldsymbol{h_*}\preceq \boldsymbol{\overline{h}}$.
    Moreover, let $\boldsymbol{\bar{\bar{z}}}$ be a fixed point of \eqref{recursion_regular_trees} with components $\bar{\bar{z}}_a^{(b)}  \in[h_a+\underline h_a^{(b)},\,h_a+\overline h_a^{(b)}]$ for every $(a,b)\in\mathcal E$. We can consider $\boldsymbol{\bar{\bar{z}}}$ as the starting point of the recursion \eqref{recursion_regular_trees} with extra boundary fields $\boldsymbol{h_*}$ with components $h_{*a}^{(b)}:= \bar{\bar{z}}_a^{(b)} - h_a \in[\underline h_a^{(b)},\,\overline h_a^{(b)}]$ for every $(a,b)\in\mathcal E$.  Then by \eqref{uniqueness_november} we have that $\boldsymbol{\bar{\bar{z}}} = \boldsymbol{\bar{z}}$. 
    
    The previous results extend to $\boldsymbol{\overline h}=+\boldsymbol{\infty}$. Indeed,
    \begin{equation}
        \boldsymbol{z}(1,+\infty) \,=\, \boldsymbol{z}(0,\boldsymbol{h_*})
    \end{equation}
    where the vector $\boldsymbol{h_*}$ has components $h_{*a}^{(b)}:=\sum_{c\in\mathcal E_a}(k_{ac}-\delta_{bc})\,\beta_{ac} \in[\underline h_a^{(b)},\overline h_a^{(b)}]$ for every $(a,b)\in\mathcal E$. Then by \eqref{uniqueness_november} it follows that $\boldsymbol{\bar{z}}(+\infty) = \boldsymbol{\bar{z}}$.
\end{proof}

\begin{remark} \label{remark different condition uniqueness}
    If conditions \eqref{condition_H_small}, \eqref{positive_boundary} hold true, condition \eqref{strict_condition_hsmall} can be replaced by a weaker one in the hypothesis of Proposition \ref{lemma_uniqueness}.
    If for every $(a,b)\in \mathcal E$ with $k_a\geq2$ there exists $s^*=s^*_{ab}\in\mathbb N$ such that 
    \begin{equation}
        z_a^{(b)}(s^*,\boldsymbol{\underline h}) \,>\, z_a^{(b)}(s^*-1,\boldsymbol{\underline h})\;,
\label{different_condition_uniqueness}
    \end{equation}
    then, by continuity of $\boldsymbol{ h_*}\mapsto  z_a^{(b)}(s,\boldsymbol{h_*})$, it is possible to find $\rho>0$ sufficiently small so that
    \begin{equation}
          z_a^{(b)}(s^*,\boldsymbol{\underline h}) \,\geq\,  z_a^{(b)}(s^*-1,\,\boldsymbol{\underline h}+\rho\,\boldsymbol{1})
\label{different_condition_uniqueness_htilde}
    \end{equation} 
    and one can adapt the arguments used in the proof of Proposition \ref{lemma_uniqueness}.
    Notice that inequality \eqref{strict_condition_hsmall} is nothing but \eqref{different_condition_uniqueness} for $s^*=1$. 
\end{remark}

\begin{proposition}[Existence and uniqueness of fixed point at high temperature]\label{prop: uniqueness for high temp}
    Suppose $\rho(\boldsymbol{M})<1$. Then there exists a unique fixed point $\boldsymbol{\bar z}$ of the recursion \eqref{recursion_regular_trees} and
    \begin{equation} 
        \boldsymbol{\bar z}(\boldsymbol{h_*})\,=\,\boldsymbol{\bar z} \quad\forall\,\boldsymbol{h_*}\in[-\infty,\infty]^{\mathcal E} \,.
    \end{equation}
\end{proposition}
\begin{proof}
     Set $\boldsymbol e(s) := \boldsymbol z(s,+\infty) - \boldsymbol z(s,-\infty)$ for every $s\in\mathbb N$. By recursion \eqref{recursion_regular_trees} we get
    \begin{equation} \left\{
    \begin{split}
        &\,e_a^{(b)}(s) \,\leq\,  \sum_{c\in\mathcal E_a} (k_{ac}-\delta_{bc})\,\tanh(\beta_{ac})\, e_c^{(a)}(s-1) \,,\quad s\geq2 \\
        &\,e_a^{(b)}(1) \,=\,  
        2 \sum_{c\in\mathcal E_a} (k_{ac}-\delta_{bc})\,\beta_{ac} \,,\quad s=1
    \end{split} \right.
    \label{recursion e(s)}
    \end{equation} 
    for every $(a,b)\in\mathcal E$, where for the first line we have linearized $F_{\beta}(x_2)-F_{\beta}(x_1) \leq \max_{x\in [x_1,x_2]}F'_{\beta}(x)\,(x_2-x_1)\leq \tanh(\beta)\,(x_2-x_1)$, while in the second line we have used $F_{\beta}(+\infty)- F_{\beta}(-\infty)=2\,\beta$. We then rewrite the first line of \eqref{recursion e(s)} as
    \begin{equation}
        \boldsymbol{e}(s) \,\preceq\, \boldsymbol{M}\, \boldsymbol{e}(s-1)
    \end{equation}
    where $\preceq$ denotes componentwise inequality and matrix $\boldsymbol{M}$ is defined in \eqref{M: definition}. Since $\boldsymbol{e}(s)$ and $\boldsymbol M$ have non-negative entries, the previous inequality can be iterated finding
    \begin{equation}
        \boldsymbol{e}(s) \,\preceq\, \boldsymbol{M}^{s-1}\, \boldsymbol{e}(1) \;.
    \end{equation}
    Therefore, choosing a suitable operator norm $\lVert\cdot\rVert$, we have
    \begin{equation}
         |\boldsymbol{e}(s)| \,\leq\,  |\boldsymbol{M}^{s-1}\,\boldsymbol{e}(1)| \,\leq\, \lVert \boldsymbol{M} ^{s-1}\rVert\, |\boldsymbol{e}(1)|\;.
    \label{eq: iterated norm}
    \end{equation}     
    Since by hypothesis  $\rho(M)< 1$, Gelfand's formula \cite{dunford1963spectral} implies the existence of $s_0\geq2$  such that $\lVert M^{s_0}\rVert <1$. Hence, from \eqref{eq: iterated norm} it follows
    \begin{equation}
         | \boldsymbol{e}(ss_0+1) |  \,\leq\, \lVert \boldsymbol{M} ^{ss_0}\rVert \,|\boldsymbol{e}(1)| \,\leq\, \lVert \boldsymbol{M} ^{s_0}\rVert^{s} \, |\boldsymbol{e}(1)| \,\xrightarrow[s\to\infty]{}\, 0 \;.
    \end{equation}
    Therefore
    \begin{equation}
        \boldsymbol{\bar{z}}(-\boldsymbol{\infty}) \,=\, \boldsymbol{\bar{z}}(+\boldsymbol{\infty})  \;.
    \end{equation}
    The uniqueness of the fixed point of \eqref{recursion_regular_trees}, reached from any initialization $\boldsymbol h_*$, then follows by Lemma \ref{corollary_monotony_H}. 
\end{proof}

\subsection{Results on trees for different external fields}
We derive consequences of Propositions \ref{lemma existence fixed point H_small H_big}, \ref{lemma_uniqueness}, \ref{prop: uniqueness for high temp}, for different choices of the external fields $h_1,\dots,h_n\,$.

\begin{corollary}\label{corollary_final_strictlypositive_fields}
    Suppose $h_1,\dots,h_n>0$.
    Then there exists a unique fixed point $\boldsymbol{\bar z}$ of the recursion \eqref{recursion_regular_trees} having all non-negative components, precisely its components satisfty $\bar z_a^{(b)}\geq h_a> 0$ for every $(a,b)\in\mathcal E$.
    Moreover
    \begin{equation} \label{zbar_lim}
        \boldsymbol{\bar z}(\boldsymbol{h_*})\,=\,\boldsymbol{\bar z} \quad\forall\,\boldsymbol{h_*}\in[0,\infty]^{\mathcal E} \,.
    \end{equation}
\end{corollary}

\begin{proof}
    It follows immediately by Proposition \ref{lemma_uniqueness} taking $\boldsymbol{\underline h}=\boldsymbol{0}\,$. 
\end{proof}

\begin{corollary}\label{corollary_final_allnonnegative_fields_one_positive}
    Suppose $h_1,\dots,h_n\geq0$ at least one of them strictly positive, and $k_1,\dots,k_n\geq2$.
    Then there exists a unique fixed point $\boldsymbol{\bar z}$ of the recursion \eqref{recursion_regular_trees} having all non-negative components.
    Moreover \eqref{zbar_lim} holds true.
\end{corollary}

The proof requires the following Lemma, which is proven in Appendix \ref{appendix: auxiliary}.
\begin{lemma}\label{lemma topology h null}
    Let $(a,b)\in\mathcal E$ and $f=1,\dots,n$. If $k_1,\dots,k_n\geq 2$, then there exists $s\in\mathbb N$ such that the tree $T_a^{(b)}(s)$ contains a vertex of class $C_f$. We denote by $s^*_{ab}(f)$ the smallest of such $s$.
\end{lemma}

\begin{proof}[Proof of Corollary \ref{corollary_final_allnonnegative_fields_one_positive}]
    Taking $\boldsymbol{\underline h}=\boldsymbol{0}\,$, the Corollary follows from Remark \ref{remark different condition uniqueness} and Proposition \ref{lemma_uniqueness}. Indeed, set $s^*_{ab}:=\min\{s^*_{ab}(f)\,|\, f=1,\dots,n,\,h_f>0\}$ for every $(a,b)\in\mathcal E$, as defined in Lemma \ref{lemma topology h null}.  The recursion \eqref{recursion_regular_trees} brings $z_a^{(b)}(s,0)=0$ for $s=1,\dots,s^*_{ab}-1$, while $z_a^{(b)}(s^*_{ab},0)>0$. Therefore condition \eqref{different_condition_uniqueness} is satisfied.
\end{proof}

\begin{corollary} \label{prop: zero fields rho(M)>1}
    Suppose $h_1=\ldots=h_n=0$.
    \begin{itemize}
        \item[i)] If $\rho(\boldsymbol{M})<1$, then $\boldsymbol 0$ is the unique fixed point of recursion \eqref{recursion_regular_trees}. Moreover
        \begin{equation}
            \boldsymbol{\bar z}(\boldsymbol {h_*}) \,=\, \boldsymbol 0 \quad\forall\,\boldsymbol {h_*}\in[-\infty,\infty]^{\mathcal E}
        \end{equation}
        \item[ii)] Assume also $k_1,\ldots,k_n\geq 2$ and $\mathcal G$ not simply cyclic. If $\rho(\boldsymbol{M})>1$, then there exist exactly 3 fixed points of the recursion \eqref{recursion_regular_trees}: $\boldsymbol{\bar z},\,\boldsymbol{0},\,-\boldsymbol{\bar z}\,$, where $\boldsymbol{\bar z}$ have all strictly positive components.     Moreover,
        \begin{equation} \label{eq: h=0 positive fixed point}
             \boldsymbol{\bar z}(\boldsymbol{h_*})\,=\,\boldsymbol{\bar z} \quad\forall\,\boldsymbol{h_*}\in(0,+\infty]^{\mathcal E} \,.
        \end{equation}
    \end{itemize}
\end{corollary}

The proof requires the following Lemma, which is proven in Appendix \ref{appendix: auxiliary}.

\begin{lemma}\label{lemma: irreducibility of M}   
    Suppose $k_1,\ldots,k_n\geq 2$ and $\mathcal G$ not simply cyclic. 
    Then the matrix $\boldsymbol M$ is irreducible.
\end{lemma}

\begin{proof}[Proof of Corollary \ref{prop: zero fields rho(M)>1}]
    Part \textit{i)} immediately follows from Proposition \ref{prop: uniqueness for high temp} observing that $\boldsymbol 0$ is always a fixed point of recursion \eqref{recursion_regular_trees} for  $h_1=\ldots=h_n=0$.
    
    Part \textit{ii)}. We claim that there exists a unique fixed point $\boldsymbol {\bar z}$ with all positive components. The result then follows by simmetry of the recursion at $h_1=\dots=h_n=0$.
    
    Now, since the matrix $\boldsymbol{M}$ is irreducible (Lemma \ref{lemma: irreducibility of M}) and $\rho(\boldsymbol{M})>1$ is its largest eigenvalue, by Perron-Frobenius theorem there exists an eigenvector $\boldsymbol{v}\succ \boldsymbol{0}$, $|\boldsymbol v|=1$, with all positive components. Hence
    \begin{equation} \label{eq:Perron-Frobenius}
        \boldsymbol M\,  \boldsymbol{v} \,=\, \rho(\boldsymbol M)\,\boldsymbol{v} \,\succ\, \boldsymbol{v} \;.
    \end{equation}
    The claim follows by Proposition \ref{lemma_uniqueness} choosing $\boldsymbol{\underline h} \,:=\, \varepsilon\, \boldsymbol{v}$ for $\varepsilon>0$ sufficiently small. Indeed condition \eqref{positive_boundary} is obviously fulfilled as  $\boldsymbol{\underline h}\succ\boldsymbol 0$ and $h_1=\ldots=h_n=0$. Condition \eqref{strict_condition_hsmall} at $h_1=\ldots=h_n=0$, 
    by expanding $F_\beta(x)$ around $x=0$, rewrites as
    \begin{equation}
         \sum_{c\in\mathcal E_a}\,(k_{ac}-\delta_{bc})\,\tanh(\beta_{ac})\;\underline h_{c}^{(a)} \,+\, O(|\boldsymbol{\underline h}|^2) \,>\, \underline h_{a}^{(b)} \quad\quad \forall \,(a,b)\in\mathcal E\,,
    \end{equation}
    which by definition \eqref{M: definition} of the matrix $\boldsymbol{M}$ is simply
    \begin{equation}
         \boldsymbol M\, \boldsymbol{\underline h} \,+\, O(|\boldsymbol{\underline h}|^2) \,\succ\, \boldsymbol{\underline h} \,.
    \end{equation}
    Dividing both sides by $\varepsilon$ we end up with $\boldsymbol M\, \boldsymbol{v} \,+\, O(\varepsilon) \,\succ\, \boldsymbol{v}\,$.
    Thanks to \eqref{eq:Perron-Frobenius} the latter inequality is verified for $\varepsilon$ sufficiently small.
\end{proof}

\section{From trees to random graphs}
\begin{proposition}
    Let $G_N=(V_N,E_N)$ be a random $\mathbf{k}$-regular graph. Fix $t\in\mathbb N$ and a vertex $i\in V_N$. Let $B_N(i,t)$ denote the ball of radius $t$ and center $i$ in the graph $G_N$.
    Then
    \begin{equation}
        \mathbb P\big(B_N(i,t)\textrm{ is a tree}\big) \,\xrightarrow[N\to\infty]{}\, 1 \;.
    \end{equation}
\label{th: tree-like}
\end{proposition}

\begin{proof}
    Let us fix a total order on the vertices of $G_N$, which follows their distance from $i$. 
    For every vertex $j$, denote by $\Gamma(j)$ the connected subgraph of $G_N$ generated by $j$ together with all the previous vertices. We have:
    \begin{equation} \label{treelike1}
        \mathbb P\big(B_N(i,t)\textrm{ is a tree}\big) \,=\, \prod_{j\in B_N(i,t)\smallsetminus\{i\}} \mathbb P\big(\Gamma(j)\textrm{ is a tree} \,\big|\, \Gamma(j-1)\textrm{ is a tree} \big) \;.
    \end{equation}
    Let $j\in B_N(i,t)\smallsetminus\{i\}$. $\Gamma(j)$ is connecteed by construction. Hence, knowing $\Gamma(j-1)$ is a tree, $\Gamma(j)$ is not tree if and only if the vertex $j$ has (at least one) extra connection to $\Gamma(j-1)$. 
    Then, since the connections of $j\in C_b$ are chosen uniformly at random among the available possibilities, we have:
    \begin{equation} \label{treelike2} \begin{split}
        &\mathbb P\big(\Gamma(j)\textrm{ is NOT a tree} \,\big|\, \Gamma(j-1)\textrm{ is a tree} \big) \,\leq\\[2pt]
        &\leq\, \sum_{c\in\mathcal E_b}k_{bc}\,\frac{\textrm{n° of vertices of class $C_c$ in }\Gamma(j-1)}{\textrm{n° of vertices of class $C_c$ in } G_N\smallsetminus\Gamma(j-1)} \\
        &\leq\, \sum_{c\in\mathcal E_b}k_{bc}\,\frac{\sum_{s=0}^{t-1} {\hat k}^s}{ \alpha_c\,N - \sum_{s=0}^{t-1} {\hat k}^s} \,=\, O(N^{-1}) 
    \end{split}\end{equation}
    setting $\hat k:=\max_{d} \sum_{d'}k_{dd'}\,$.
    Therefore, plugging \eqref{treelike2} into \eqref{treelike1} we find:
    \begin{equation}
        \mathbb P\big(B_N(i,t)\textrm{ is a tree}\big) \,=\, \prod_{j\in B_N(i,t)\smallsetminus\{i\}} \big(1-O(N^{-1})\big) \,=\, 1 + O(N^{-1})
    \end{equation}
    since the cardinality of $B_N(i,t)$ is bounded by $\sum_{s=0}^t {\hat k}^s$ independently of $N$.
\end{proof}

\begin{remark}
    If the ball $B_N(i,t)$ is a tree, then it is isomorphic to the $\mathbf{k}$-regular tree $T_a(t,\mathbf{k})$, where $C_a$ is the class of vertex $i$.
\end{remark}

\subsection{Proof of Theorems \ref{main:magnetization}, \ref{main: high temperatures}, and \ref{main: phase transition}}

\begin{proof}[Proof of Theorem \ref{main:magnetization}]
   Let us restrict the Ising model \eqref{Hamiltonian} to the ball $B_N(i,t)$, choosing free and positive boundary conditions: they correspond respectively to no extra fields and infinite fields attached to the vertices at distance $t$ from $i$. Let us denote by $\langle\sigma_i\rangle_{B_N(i,t)}^0$ and $\langle\sigma_i\rangle_{B_N(i,t)}^+$ the corresponding equilibrium magnetization of spin $\sigma_i$.
   By GKS inequalities, applying the same method (removal of edges and addition of fields) described for trees in the proof of Remark \ref{remark_proof_GKS} (see Appendix), we have:
   \begin{equation} \label{GKS_ball}
   \langle\sigma_i\rangle_{B_N(i,t)}^0 \;\;\leq \;\; \langle\sigma_i\rangle_{G_N} \;\;\leq \;\; \langle\sigma_i\rangle_{B_N(i,t)}^+ \,.
   \end{equation}
   By Proposition \ref{th: tree-like} the ball $B_N(i,t)$ is isomorphic to the $\mathbf{k}$-regular tree $T_a(t)$ with probability convergent to $1$ as $N\rightarrow \infty$.
   Let us denote by $\langle\sigma_i\rangle_{T_a(t)}^0$, $\langle\sigma_i\rangle_{T_a(t)}^+$ the root magnetisations in the Ising model \eqref{perturbed hamiltonian} on the tree $T=T_a(t)$, with extra fields at the boundary respectively $h_*=0$, $h_*=+\infty$. 
   By Corollary \ref{corollary_final_allnonnegative_fields_one_positive} we have:
   \begin{equation} \label{+0convergence}
    \langle\sigma_i\rangle_{T_a(t)}^+ \,\searrow \, \bar{m}_a  \ ,\quad\  \langle\sigma_i\rangle_{T_a(t)}^0 \,\nearrow\, \bar{m}_a   \qquad\textrm{as }t\to\infty
   \end{equation}
   where 
   \begin{equation}
    \bar{m}_a  \,:=\, \tanh\bigg( h_a + \sum_{c\in \mathcal E_a} k_{ac}\, F_{\beta_{ac}}(\bar{z}_c^{(a)}) \bigg)
   \end{equation}
   and $\boldsymbol{\bar z}=(\bar{z}_b^{(c)})_{(b,c)\in \mathcal E}$ is the unique fixed point of the recursion \eqref{recursion_regular_trees} with all non-negative entries.
   Now let $\epsilon,\delta>0$. Using inequalities \eqref{GKS_ball} we can write 
   \begin{equation} \begin{split}
        \langle\sigma_i\rangle_{G_N} - \bar{m}_a \, &\leq\, \langle\sigma_i\rangle_{B_N(i,t)}^+ -  \bar{m}_a  \\[2pt]
       &\leq\; \Big|\langle\sigma_i\rangle_{B_N(i,t)}^+-\langle\sigma_i\rangle_{T_a(t)}^+\Big| \,+\, \Big|\langle\sigma_i\rangle_{T_a(t)}^+  - \bar{m}_a \Big|
   \end{split} \end{equation}
   and
    \begin{equation} \begin{split}
       \bar{m}_a - \langle\sigma_i\rangle_{G_N}  \, &\leq \, \bar{m}_a - \langle\sigma_i\rangle_{B_N(i,t)}^0  \\[2pt]
       &\leq\, \Big|\bar{m}_a - \langle\sigma_i\rangle_{T_a(t)}^0\Big| \,+\, \Big|\langle\sigma_i\rangle_{T_a(t)}^0- \langle\sigma_i\rangle_{B_N(i,t)}^0\Big| \;.
   \end{split}\end{equation}
   By convergence \eqref{+0convergence} and Proposition \ref{th: tree-like}, there exist $t_\epsilon$, $N_{\delta,t_\epsilon}$ finite such that 
   \begin{equation}
       \Big|\langle\sigma_i\rangle_{T_a(t)}^+  -  \bar{m}_a\Big|   \,<\, \epsilon  \;,\quad \Big|\bar{m}_a - \langle\sigma_i\rangle_{T_a(t)}^0\Big|  \,<\,\epsilon \quad \forall\, t\geq t_\epsilon 
   \end{equation}
   \begin{equation}
       \Pprob\bigg(\langle\sigma_i\rangle_{B_N(i,t_\epsilon)}^+\neq\langle\sigma_i\rangle_{T_a(t_\epsilon)}^+\ \textrm{ or }\ \langle\sigma_i\rangle_{B_N(i,t_\epsilon)}^0\neq\langle\sigma_i\rangle_{T_a(t_\epsilon)}^0 \bigg) \,<\,\delta  \quad \forall \,N\geq N_{\delta,t_\epsilon} \;.
   \end{equation}
   Therefore combining the previous inequalities,
   \begin{equation}
   \Pprob\bigg(\big| \langle\sigma_i\rangle_{G_N} -\; \bar{m}_a\big| \geq \epsilon \bigg) \,<\, \delta  \quad \forall \, N>N_{\delta,t_\epsilon} \,.
   \end{equation}
    As $\langle\sigma_i\rangle_{G_N} -\; \bar{m}_a$ is bounded, convergence in probability implies $L^r$-norm convergence, and we have proved \eqref{eq: main magnetization}.\\
    
    Now, let us restrict the Ising model \eqref{Hamiltonian} to the ball $B_N(ij,t)$ of radius $t$ centered in the edge $ij\in E_N$, i.e., the subgraph of $G_N$ generated by the vertices at distance $\leq t$ from either $i$ or $j$. By the GKS inequalities \eqref{GKS_ij_h}, \eqref{GKS_ij_beta}, 
    \begin{equation}
        \langle\sigma_i\sigma_j\rangle_{B_N(ij,t)}^0 \quad \leq \quad \langle\sigma_i\sigma_j\rangle_{G_N} \quad \leq \quad \langle\sigma_i\sigma_j\rangle_{B_N(ij,t)}^+
    \label{eq: two point correlation ball}
    \end{equation}
    Superscripts $0$ and $+$ refer, as usual, to free and plus boundary conditions respectively.
    By Propostion \ref{th: tree-like},
    \begin{equation}
    \mathbb P\big(B_N(ij,t) \textrm{ is a tree}\big) \,\xrightarrow[N\to\infty]{}\, 1
    \end{equation}
    and this tree is necessarely the $\mathbf{k}$-regular tree $T_{aa'}(t)$ of depth $t$ rooted in an edge connecting two vertices of classes $a,a'$  (each of them generates a first generation and so on up to generation $t$).
    By an application of Lemma \ref{lemma general two points correlation} and Corollary \ref{corollary_final_allnonnegative_fields_one_positive}, we have
     \begin{equation}
         \langle\sigma_i\sigma_j\rangle_{T_{aa'}(t)}^0  \nearrow \;\bar{\gamma}_{aa'} \;, \qquad  \langle\sigma_i\sigma_j\rangle_{T_{aa'}(t)}^+ \searrow \;\bar{\gamma}_{aa'} \qquad \textrm{as }t\rightarrow \infty\,,
     \label{converg_eq}
     \end{equation}
     where $\bar{\gamma}_{aa'} = \tanh\Big(  \beta_{aa'} + \tanh^{-1}\big( \tanh(\bar{z}_a^{(a')})\tanh(\bar{z}_{a'}^{(a)}) \big) \Big) $.
     Indeed \eqref{eq general two points} is monotonic non-decreasing with respect to the product $m_a^{(a')} m_{a'}^{(a)}$. Now, following the same procedure already explained for the magnetisation we obtain the convergence \eqref{eq: main two point correlation}.
\end{proof}

\begin{proof}[Proof of Theorem \ref{main: high temperatures}]
    Let us restrict the Ising model \eqref{Hamiltonian} to the ball $B_N (i, t)$, 
    choosing respectively negative and positive boundary conditions, i.e., attaching $\pm\infty$ fields to the vertices at distance $t$ from $i$ in $G_N$. FKG inequalities (see \eqref{FKG} respectively addition and removal of fields) always guarantee that:
    \begin{equation} \label{FKG_ball}
        \langle\sigma_i\rangle_{B_N(i,t)}^- \;\;\leq \;\; \langle\sigma_i\rangle_{G_N} \;\;\leq \;\; \langle\sigma_i\rangle_{B_N(i,t)}^+ \,.
    \end{equation}
   On the other hand, let us denote by $\langle\sigma_i\rangle_{T_a(t)}^{\pm}$ the root magnetisation of the Ising model \eqref{perturbed hamiltonian} on the tree $T=T_a(t)$, with extra field at the boundary respectively $h_*=\pm\infty$. 
   For $\rho(\boldsymbol M)<1$, by Corollary \ref{prop: uniqueness for high temp} we have:
   \begin{equation} \label{+0convergence}
    \langle\sigma_i\rangle_{T_a(t)}^- \,\nearrow \, \bar{m}_a  \ ,\quad\  \langle\sigma_i\rangle_{T_a(t)}^+ \,\searrow\, \bar{m}_a   \qquad\textrm{as }t\to\infty
   \end{equation}
   where 
   \begin{equation}
    \bar{m}_a  \,:=\, \tanh\bigg( h_a + \sum_{c\in \mathcal E_a} k_{ac}\, F_{\beta_{ac}}(\bar{z}_c^{(a)}) \bigg)
   \end{equation}
   and $\boldsymbol{\bar z}=(\bar{z}_b^{(c)})_{(b,c)\in \mathcal E}$ is the unique fixed point of the recursion \eqref{recursion_regular_trees}.
   The proof now proceeds in the same way as the proof of Theorem \ref{main:magnetization}, obtaining the convergence \eqref{eq: main magnetization}.
\end{proof}

\begin{proof}[Proof of Theorem \ref{main: phase transition}]
    %
    %
    By Theorem \ref{main:magnetization}, the spontaneous magnetisation \eqref{eq: spontaneous} is given by the following expression
    \begin{equation} \label{eq: spontaneous_expression}
        S_a(\boldsymbol \beta) \,=\, \lim_{h\to0+} \tanh\Big( h \,+\, \sum_{c\in\mathcal E_a} k_{ac}\, F_{\beta_{ac}}\big(\bar{z}_c^{(a)}\big) \Big)
    \end{equation} 
     where for every $h>0$, $\boldsymbol{\bar z}=\boldsymbol{\bar z}(\boldsymbol{\beta},h)$ is the only solution with all positive components of the fixed point equation
     \begin{equation}
             \bar z_a^{(b)} \,=\, h + \sum_{c\in \mathcal E_a} (k_{ac}-\delta_{bc})\, F_{\beta_{ac}}\big(\bar z_c^{(a)}\big)\,, \quad (a,b)\in\mathcal E \,.
    \label{fixed point eq with field}
     \end{equation}
    The limit $\lim_{h\to 0^+}\boldsymbol{\bar z}(\boldsymbol{\beta},h)$ is reached with decreasing monotonicity (Lemma \ref{corollary_monotony_H}), and by continuity of the function $F_\beta(x)$ it is a solution of 
     \begin{equation}
             \tilde z_a^{(b)} \,=\, \sum_{c\in \mathcal E_a} (k_{ac}-\delta_{bc})\, F_{\beta_{ac}}\big( \tilde z_c^{(a)}\big)\,, \quad (a,b)\in\mathcal E\,.
    \label{fixed point eq no field}
     \end{equation}
     
    If $\rho(\boldsymbol M)<1$, we know that $\boldsymbol 0$ is the unique solution of fixed point equation \eqref{fixed point eq no field} (Corollary \ref{prop: zero fields rho(M)>1}). Therefore $\lim_{h\to 0+}\boldsymbol{\bar z}(\boldsymbol\beta,h)=\boldsymbol{0}$. It follows $S_a(\boldsymbol\beta)=0$.
    
    On the contrary, if $\rho(\boldsymbol M)> 1$ equation \eqref{fixed point eq no field} has also a (unique) solution $\boldsymbol{\bar z^+}=\boldsymbol{\bar z^+}(\boldsymbol{\beta})\succ \boldsymbol{0}$ with all positive components (Corollary \ref{prop: zero fields rho(M)>1}). 
    We claim that
    \begin{equation} \label{claim+}
        \boldsymbol{\bar z}(\boldsymbol\beta,h) \,\succeq\, \boldsymbol{\bar z^+}(\boldsymbol\beta) \quad\forall\,h>0 \;.
    \end{equation}
    Then letting $h\to0+$
    \begin{equation}
        \lim_{h\to0+}\boldsymbol{\bar z}(\boldsymbol\beta,h) \,\succeq\, \boldsymbol{\bar z^+}(\boldsymbol\beta) 
    \end{equation}
    and by uniqueness it follows that $\lim_{h\to0+}\boldsymbol{\bar z}(\boldsymbol\beta,h) \,=\, \boldsymbol{\bar z^+}(\boldsymbol\beta)$. Replacing the limit in \eqref{eq: spontaneous_expression} concludes the proof.
    In order to show the validity of claim \eqref{claim+}, observe that $\boldsymbol{\bar z}(\boldsymbol\beta,h)$ for $h>0$ is reached as $s\to\infty$ by the following recursion:
    \begin{equation} \left\{
     \begin{split}
        &\,z_a^{(b)} (s) \,=\, h \,+\, \sum_{c\in\mathcal E_a}\, (k_{ac}-\delta_{bc})\,F_{\beta_{ac}}\big(z_c^{(a)} (s-1)\big) \,,\quad s\geq1 \\
        &\,z_a^{(b)} (0) \,=\, \bar z_a^{(b)+}(\boldsymbol \beta)>0 \,,\quad s=0 
    \end{split} \right.
    \label{recursion h>0}
    \end{equation}
    for all $(a,b)\in\mathcal E$ (Corollary \ref{corollary_final_strictlypositive_fields}). Then, using positivity of $h$ and monotonicity of the function $F_\beta(x)$, it is easy to prove \eqref{claim+} by induction on $s$.   
\end{proof}

\subsection{Proof of Theorem \ref{main:pressure}}
\begin{proof}[Proof of Theorem \ref{main:pressure}]
    The proof consists in different steps.
    \begin{enumerate}[]
        \item Show that at $\boldsymbol{\beta}=\boldsymbol{0}$ we have $p_{G_N}(\boldsymbol{0},\boldsymbol{h})= p(\boldsymbol{0},\boldsymbol{h})$ for every $N$.
      
        \item Show that for every $(a,a')\in\mathcal E$
        \begin{equation}
             \frac{\partial p_{G_N}(\boldsymbol{\beta},\boldsymbol{h})}{\partial \beta_{aa'}} \,=\, \frac{1}{N} \sum_{\substack{ij\in E_N:\\ i\in C_a, j\in C_{a'}}} \!\langle\sigma_i \sigma_j\rangle_{G_N}   \ \ \xrightarrow[N\rightarrow \infty]{L^r(\Pprob)}\ \ \frac{\alpha_a\, k_{aa'}}{2^{\delta_{aa'}}}\, \langle\sigma_1 \sigma_2\rangle_{aa'} 
        \label{eq_edge_covergence}
        \end{equation}
        where $\delta_{aa'}$ is the Kronecker delta and $\langle\cdot\rangle_{aa'}$ is the Gibbs measure associated to the following two spins Hamiltonian
        \begin{equation}
            H_{aa'}(\sigma_1,\sigma_2) \,=\, -\,\bar{z}_a^{(a')}\,\sigma_1 \,-\, \bar{z}_{a'}^{(a)}\,\sigma_2 \,-\, \beta_{aa'}\,\sigma_1\sigma_2\,,\quad(\sigma_1,\sigma_2)\in\{\pm1\}^2\,.
        \end{equation}

        \item Show that for every $(a,a')\in\mathcal E$ also
        \begin{equation}
             \frac{\partial p(\boldsymbol{\beta},\boldsymbol{h})}{\partial \beta_{aa'}} \,=\,   \frac{\alpha_a\,k_{aa'}}{2^{\delta_{aa'}}}\,\langle\sigma_1 \sigma_2\rangle_{aa'}
        \label{derivative_functional}
        \end{equation}

        \item The previous steps entail the convergence of the derivatives at any $\boldsymbol{\beta}$ and the equivalence of the pressures at $\boldsymbol{\beta}=\boldsymbol{0}$. Finally use the fundamental theorem of calculus and dominated convergence to prove (\ref{eq_free_en_convergence}).
    \end{enumerate}

    \begin{enumerate}[label=\textit{\arabic*}.]
        \item For the proof of the first point it suffices to compute the partition function in the non-interacting case:
        \begin{equation}\begin{split}
            Z_{G_N}(\boldsymbol{0},\boldsymbol{h}) \,&=\, \sum_{\boldsymbol{\sigma}\in\{-1,1\}^{V_N}}e^{\sum_{a=1}^nh_a\sum_{i\in C_a}\sigma_i} \,=\,\prod_{a=1}^n \prod_{i\in C_a} \sum_{\sigma_i=\pm1}e^{h_a\sigma_i}\\
            &=\, \prod_{a=1}^n \big(2\cosh{(h_a)}\big)^{N\alpha_a}
        \end{split}\end{equation}
        then
        \begin{equation}
            p_{G_N}(\boldsymbol{0},\boldsymbol{h})\,=\, N^{-1}\log Z_{G_N}(\boldsymbol{0},\boldsymbol{h}) \,=\, \log2 \,+\, \sum_{a=1}^n \alpha_a \log\cosh(h_a) \,.
        \end{equation}
        On the other hand subsituting $\boldsymbol{\beta}=\boldsymbol{0}$ in the expression \eqref{free_energy} one can see that also
        \begin{equation}
            p(\boldsymbol{0},\boldsymbol{h}) \,=\, \log2 + \sum_{a=1}^n \alpha_a \log\cosh(h_a) \,.
        \end{equation}

        \item
        A standard computation shows the first identity in \eqref{eq_edge_covergence}. 
        On the other hand we can compute
        \begin{equation}
        \begin{split}
            \langle\sigma_1 \sigma_2\rangle_{aa'} &=\,  \frac{e^{\beta_{aa'}+\bar{z}_a^{(a')}+\bar{z}_{a'}^{(a)}} + e^{\beta_{aa'}-\bar{z}_a^{(a')}-\bar{z}_{a'}^{(a)}} - e^{-\beta_{aa'}+\bar{z}_a^{(a')}-\bar{z}_{a'}^{(a)}} - e^{-\beta_{aa'}-\bar{z}_a^{(a')}+\bar{z}_{a'}^{(a)}} }{ e^{\beta_{aa'}+\bar{z}_a^{(a')}+\bar{z}_{a'}^{(a)}} + e^{\beta_{aa'}-\bar{z}_a^{(a')}-\bar{z}_{a'}^{(a)}} + e^{-\beta_{aa'}+\bar{z}_a^{(a')}-\bar{z}_{a'}^{(a)}} + e^{-\beta_{aa'}-\bar{z}_a^{(a')}+\bar{z}_{a'}^{(a)}} } \\
            &=\, \frac{\tanh(\beta_{aa'}) + \tanh(\bar{z}_a^{(a')}) \tanh(\bar{z}_{a'}^{(a)}) }{1+\tanh(\beta_{aa'}) \tanh(\bar{z}_a^{(a')}) \tanh(\bar{z}_{a'}^{(a)}) }\\
            &= \tanh\bigg(  \beta_{aa'} + \tanh^{-1}\bigg( \tanh(\bar{z}_a^{(a')})\tanh(\bar{z}_{a'}^{(a)}) \bigg) \bigg) 
        \label{one_edge}
        \end{split}
        \end{equation}
        where in the last step we used $\tanh(x+y)=\frac{\tanh(x)+\tanh(y)}{1+\tanh(x)\tanh(y)}$.\\
        Therefore the convergence in \eqref{eq_edge_covergence} follows from convergence \eqref{eq: main two point correlation} reasoning as in Remark \ref{remark:convergence_average}.

        \item The whole proof of the third point consists in the computation of $\frac{\partial p}{\partial \beta_{bb'}}(\boldsymbol{\beta},\boldsymbol{h})$ from expression \eqref{free_energy}. 
        Let us first consider a generic vector $\boldsymbol{x} = \{ x_j\in \mathbb{R}\}_{j\in J}$ and a parameter $h\in\mathbb R$. Set
        \begin{align}
            K^{\pm}(\boldsymbol{x},h) \,&:=\, e^h \prod_{j\in J} (1+ x_j) \,\pm\, e^{-h} \prod_{j\in J} (1- x_j) 
            \label{eq:K_def}\\
            G(\boldsymbol{x},h) \,&:=\, \frac{K^{-}(\boldsymbol{x},h) }{K^{+}(\boldsymbol{x},h) }  \;.
            \label{def_G_x}
        \end{align}
        For $i\in J$ consider a vector with one component less $\boldsymbol{x}_{-i} = \{ x_j \}_{j\in J\smallsetminus\{i\}}\,$. We have:
        \begin{equation}
            \frac{K^{-}(\boldsymbol{x}_{-i},h) }{K^+(\boldsymbol{x},h)} \,=\, \frac{K^{-}(\boldsymbol{x}_{-i},h)/K^{+}(\boldsymbol{x}_{-i},h) }{K^+(\boldsymbol{x},h)/K^{+}(\boldsymbol{x}_{-i},h)} \,=\, \frac{G(\boldsymbol{x}_{-i},h)}{1+x_i\,G(\boldsymbol{x}_{-i},h)}
        \label{property_algebra}
        \end{equation}
        using the identity $K^+(\boldsymbol{x},h) - K^+(\boldsymbol{x}_{-i},h) = x_i\, K^-(\boldsymbol{x}_{-i},h)$ which is easy to check.
        On the other hand, the function $G$ rewrites as:
        \begin{equation}
        \begin{split}
            G(\boldsymbol{x},h) \,&=\, \frac{e^{2\log \big(e^{h} \prod_{j\in J} (\frac{1+ x_j}{1- x_j})^{\frac{1}{2}}\big)} - 1 }{e^{ 2\log\big( e^{h}\prod_{j\in J} (\frac{1+ x_j}{1- x_j})^{\frac{1}{2}}\big)} + 1} 
            \,=\, \tanh\,\log\bigg(e^h \prod_{j\in J} \bigg(\frac{1+ x_j}{1- x_j}\bigg)^\frac{1}{2} \bigg) \\
            &=\,  \tanh\bigg( h + \sum_{j\in J} \tanh^{-1}( x_j) \bigg) \;.
        \label{G_x}
        \end{split}
        \end{equation}
        Now, let us start computing the derivatives of $p$. For every $(b,b')\in \mathcal E$, $b\leq b'$, we have
        \begin{equation} \label{eq:derivative_p} \begin{split}
            \frac{\partial p}{\partial \beta_{bb'}} \,&=\,
            \frac{\partial \theta_{bb'}}{\partial \beta_{bb'}} \, \frac{\partial p}{\partial \theta_{bb'}} \,=\, (1-\theta_{bb'}^2)\, \frac{\partial p}{\partial \theta_{bb'}} \,=\, \frac{1-\theta_{bb'}^2}{2}\, \sum_{a\leq n} \alpha_a\,\frac{\partial f_a}{\partial \theta_{bb'}}  
        \end{split} \end{equation}
        where we set
        \begin{equation}
        \begin{split}
            f_a := & - \sum_{a'\in \mathcal E_a} k_{aa'} \bigg( \frac{\log(1-\theta_{aa'}^2)}{2} \,+\, \log\big(1+\theta_{aa'}\,\bar{m}_{a}^{(a')} \bar{m}_{a'}^{(a)}\big) \bigg) \;+  \\
            & +\,2 \log\bigg( e^{h_a} \prod_{a'\in \mathcal E_a} \big(1+\theta_{aa'} \,\bar{m}_{a'}^{(a)}\big)^{k_{aa'}} +\, e^{-h_a} \prod_{a'\in \mathcal E_a} \big(1-\theta_{aa'}\, \bar{m}_{a'}^{(a)}\big)^{k_{aa'}} \bigg) \,.
        \label{derivatives_freeen}
        \end{split}
        \end{equation}
        Notice that $\boldsymbol{\bar{m}}$ depends on $\boldsymbol{\theta}$ and it is differentiable by the implicit function theorem. Then for every $a\leq n$ we can write
        \begin{equation}
            \frac{\partial f_a}{\partial \theta_{bb'}} \,=\, \frac{\partial f_a}{\partial^{explicit} \theta_{bb'}} \,+ \sum_{(c,c')\in\mathcal E} \frac{\partial f_a}{\partial \bar m_{c'}^{(c)}}\,\frac{\partial \bar m_{c'}^{(c)}}{\partial \theta_{bb'}}
        \end{equation}
        where the first derivative on the r.h.s. is taken with respect to the explicit contribution of $\theta_{bb'}$.
        Now, let us introduce the following vector
        \begin{equation}
        \boldsymbol{x}^{(a)} :=\, \big(\underbrace{\theta_{a1}\,\bar{m}_1^{(a)}, \dots, \theta_{a1}\,\bar{m}_1^{(a)}}_{k_{a1}\textrm{ times}},\, \underbrace{\theta_{a2}\,\bar{m}_2^{(a)},\dots,\theta_{a2}\,\bar{m}_2^{(a)}}_{k_{a2}\textrm{ times}},\, \dots\dots,\, \underbrace{\theta_{an}\,\bar{m}_n^{(a)},\dots,\theta_{an}\,\bar{m}_n^{(a)}}_{k_{an}\textrm{ times}}\big)
        \end{equation}
        where each element $\theta_{aa'}\,\bar{m}_{a'}^{(a)}$ is repeated $k_{aa'}$ times.  
        For every $a'\in\mathcal E_a$, let us also introduce the vector $\boldsymbol{x}^{(a)}_{-a'}$ obtained from $\boldsymbol{x}^{(a)}$ removing   only one component of the type $\theta_{aa'}\,\bar{m}_{a'}^{(a)}$.
        Then, recalling also $\theta_{bb'} \equiv \theta_{b'b}$, we compute:
        \begin{equation} \begin{split}
            \frac{\partial f_a}{\partial^{explicit} \theta_{bb'}} \,= &  -\sum_{a'\in \mathcal E_a} \delta_{\{a,a'\},\{b,b'\}}\, k_{aa'}\bigg( \frac{-\theta_{aa'}}{1-\theta_{aa'}^2} + \frac{\bar{m}_a^{(a')}\bar{m}_{a'}^{(a)}}{1+\theta_{aa'}\,\bar{m}_{a}^{(a')} \bar{m}_{a'}^{(a)}} \bigg)  \;+\\
            &+\, 2 \sum_{a'\in \mathcal E_a} \delta_{\{a,a'\},\{b,b'\}}\,k_{aa'}\,\bar m_{a'}^{(a)}\,\frac{K^-(\boldsymbol{x}^{(a)}_{-a'}\,,h_a)}{K^+(\boldsymbol{x}^{(a)},h_a)}
        \label{derivative_explicit1}
        \end{split}\end{equation}
        where $\delta_{\{a,a'\},\{b,b'\}}=(\delta_{a,b}\,\delta_{a',b'}+\delta_{a,b'}\,\delta_{a',b})/2^{\delta_{bb'}}$.
        Applying the identity \eqref{property_algebra} we have
        \begin{equation}
            \frac{K^-(\boldsymbol{x}^{(a)}_{-a'}\,,h_a)}{K^+(\boldsymbol{x}^{(a)},h_a)} \,=\, \frac{G(\boldsymbol{x}^{(a)}_{-a'}\,,h_a)}{1+ \theta_{aa'}\,\bar m_{a'}^{(a)}\,G(\boldsymbol{x}^{(a)}_{-a'}\,,h_a)}
        \end{equation}
        and by \eqref{G_x}, since $\tanh^{-1}(\boldsymbol{\bar m})$ is a fixed point of recursion \eqref{recursion_regular_trees}, we find
        \begin{equation}
            G(\boldsymbol{x}^{(a)}_{-a'}\,,h_a) \,=\, \tanh\bigg(h_a+\sum_{c\in\mathcal E_a} (k_{ac}-\delta_{a'c})\,\tanh^{-1}\big(\theta_{ac}\,\bar m_a^{(c)}\big)\bigg) \,=\, \bar m_a^{(a')} \;.
        \end{equation}
        Plugging the previous identities into \eqref{derivative_explicit1} and rearranging terms we obtain:
        \begin{equation} \begin{split}
            \frac{\partial f_a}{\partial^{explicit} \theta_{bb'}} 
            \,=\, \sum_{a'\in \mathcal E_a} \delta_{\{a,a'\},\{b,b'\}}\, k_{aa'}\, \bigg(\frac{\theta_{aa'}}{1-\theta_{aa'}^2} \,+\, \frac{\bar{m}_a^{(a')}\bar{m}_{a'}^{(a)}}{1+\theta_{aa'}\,\bar{m}_{a}^{(a')} \bar{m}_{a'}^{(a)}} \bigg) \;. \label{eq:derivative_explicit2}
        \end{split}\end{equation}
        Now, for every $(c,c')\in\mathcal E$ let us compute
        \begin{equation} \begin{split}
            \frac{\partial f_a}{\partial\bar m_{c'}^{(c)}} \,&=\, -\sum_{a'\in \mathcal E_a} k_{aa'}\, \frac{\theta_{aa'}\Big(\bar{m}_a^{(a')}\delta_{(a,a'),(c',c)}\,+\,\bar{m}_{a'}^{(a)}\delta_{(a,a'),(c,c')}\Big)}{1+\theta_{aa'}\,\bar{m}_{a}^{(a')} \bar{m}_{a'}^{(a)}}  \;+\\
            &\quad \ +\, 2 \sum_{a'\in \mathcal E_a} \delta_{(a,a'),(c,c')}\,k_{aa'}\,\theta_{aa'}\,\frac{K^-(\boldsymbol{x}^{(a)}_{-a'}\,,h_a)}{K^+(\boldsymbol{x}^{(a)},h_a)}
        \end{split}\label{eq:derivative_implicit1}\end{equation}
        where $\delta_{(a,a'),(c,c')}=\delta_{a,c}\,\delta_{a',c'}$.
        As already seen $ \frac{K^-(\boldsymbol{x}^{(a)}_{-a'}\,,h_a)}{K^+(\boldsymbol{x}^{(a)},h_a)} = \frac{\bar m_a^{(a')}}{1+\theta_{aa'}\,\bar m_{a'}^{(a)}\,\bar m_a^{(a')}}$, hence we find
        \begin{equation} \begin{split}
            \frac{\partial f_a}{\partial\bar m_{c'}^{(c)}} 
            \,=\, 0 \;.
        \end{split}\label{eq:derivative_implicit2}\end{equation}
        Therefore
        \begin{equation}
            \frac{\partial f_a}{\partial \theta_{bb'}} \,=\, \frac{\partial f_a}{\partial^{explicit} \theta_{bb'}} \;.
        \end{equation}
        Finally, plugging expression \eqref{eq:derivative_explicit2} into \eqref{eq:derivative_p} and using the identity $\alpha_bk_{bb'}=\alpha_{b'}k_{b'b}$ we obtain:
        \begin{equation}
        \begin{split}
            \frac{\partial p}{\partial \beta_{bb'}} \,&= \, \frac{1-\theta_{bb'}^2}{2} \sum_{(a,a')\in\mathcal E} \delta_{\{a,a'\},\{b,b'\}}\,\alpha_a\,k_{aa'}\,\bigg( \frac{\theta_{aa'}}{1-\theta_{aa'}^2} + \frac{\bar{m}_a^{(a')}\bar{m}_{a'}^{(a)}}{1+\theta_{aa'}\,\bar m_{a}^{(a')}\,\bar m_{a'}^{(a)}} \bigg) \\
            &=\,\frac{\alpha_b\,k_{bb'}}{2^{\delta_{bb'}}}\;\frac{\theta_{bb'}+\bar{m}_b^{(b')}\bar{m}_{b'}^{(b)} }{1+\theta_{bb'}\,\bar{m}_b^{(b')}\bar{m}_{b'}^{(b)}}\;,
        \end{split}
        \end{equation}
        which proves \eqref{derivative_functional} by noticing again that $\frac{\theta_{bb'}+\bar{m}_b^{(b')}\bar{m}_{b'}^{(b)} }{1+\theta_{bb'}\,\bar{m}_b^{(b')}\bar{m}_{b'}^{(b)}} \,=\, \langle\sigma_1 \sigma_2\rangle_{bb'}\,$.

        \item 
        Using the Fundamental Theorem of Calculus and point 1. we write: 
        \begin{equation}
        \begin{split}
            &\big\lVert p_{G_N}(\boldsymbol{\beta},\boldsymbol{h}) -p(\boldsymbol{\beta},\boldsymbol{h}) \big\rVert_{L^r(\mathbb P)} =\\[2pt]  
            &=\,\bigg\lVert \int_0^1 \bigg( \frac{ d p_{G_N}(t\boldsymbol{\beta},\boldsymbol{h})}{d t} -  \frac{ d p(t\boldsymbol{\beta},\boldsymbol{h})}{d t}\bigg) \,dt \, \bigg\rVert_{L^r(\mathbb P)}  \\[2pt]
            &\leq\, \int_0^1 \sum_{(a,a')\in \mathcal E} \beta_{aa'}\, \bigg\lVert\frac{\partial p_{G_N}}{\partial \beta_{aa'}}(t\boldsymbol{\beta},\boldsymbol{h}) - \frac{\partial p}{\partial \beta_{aa'}} (t\boldsymbol{\beta},\boldsymbol{h})\bigg\rVert_{L^r(\mathbb P)} dt\;.   
        \end{split}
        \end{equation}  
        The latter quantity vanishes as $N\to\infty$ by points 2., 3. and dominated convergence, completing the proof.
    \end{enumerate}    
\end{proof}

\begin{remark}
        One can easily verify that
        \begin{equation}
            \frac{\partial p_N(\boldsymbol{\beta},\boldsymbol{h})}{\partial h_a} \,=\, \frac{1}{N} \sum_{\substack{i\in V_N:\\i\in C_a}} \langle \sigma_i\rangle_{G_N} \ \xrightarrow[N\rightarrow \infty]{L^r(\Pprob)}\ \frac{\partial p(\boldsymbol{\beta},\boldsymbol{h})}{\partial h_a} \,.
        \end{equation}
\end{remark}

\appendix
\section{Auxiliary results} \label{appendix: auxiliary}
In this Appendix we prove some auxiliary results used in Section \ref{sec:trees}. 

\begin{proof}[Proof of Lemma \ref{lemma_general_tree_recirsion}]
    Denote by $\boldsymbol{\sigma}_{T_w}$ the spin configuration on the subtree $T_w$ and by $H_{T_w}(\boldsymbol{\sigma}_{T_w}) = -\sum_{(i,j)\in E(T_w)} \beta_{ij} \sigma_i\sigma_j - \sum_{i\in V(T_w)} h_i \sigma_i $ the corresponding Hamiltonian. 
    We can rewrite $H_{T_w}(\boldsymbol{\sigma}_{T_w})$ as follows: 
    \begin{equation}
        H_{T_w}(\boldsymbol{\sigma}_{T_w}) \,=\,   -h_w\sigma_w \,-\, \sum_{i=1}^{K_w} \beta_{wu_i}\sigma_{u_i}\sigma_w \,+\, \sum_{i=1}^{K_w} H_{T_{u_i}}(\boldsymbol{\sigma}_{T_{u_i}}) \;.
    \end{equation}
    Setting $S_{w,T_w}(\pm1) := \mu_{T_w}(\sigma_w=\pm1)$ and $Z_{T_w}:= \sum_{\boldsymbol{\sigma}_{T_w}} e^{H_{T_w}(\boldsymbol{\sigma}_{T_w})}$, we have
    \begin{equation} \begin{split}
     S_{w,T_w}(\xi) \,&=\, \frac{1}{Z_{T_w}}\, \sum_{\boldsymbol{\sigma}_{T_w}:\sigma_w=\xi} e^{-H_{T_w}(\boldsymbol{\sigma}_{T_w})}  \\
     &=\, \frac{e^{h_w\xi}}{Z_{T_w}}\, \prod_{i=1}^{K_w} \;\sum_{\boldsymbol{\sigma}_{T_{u_i}}} e^{\beta_{wu_i} \sigma_{u_i}\xi} \;\;e^{-H_{T_{u_i}}(\boldsymbol{\sigma}_{T_{u_i}})}  \\
     &=\, \frac{e^{h_w\xi}}{Z_{T_w}}\, \prod_{i=1}^{K_w} \;\bigg(  e^{\beta_{wu_i} \xi} \sum_{\boldsymbol{\sigma}_{T_{u_i}}: \sigma_{u_i}=1} e^{-H_{T_{u_i}}(\boldsymbol{\sigma}_{T_{u_i}})}  \,+\,  e^{-\beta_{wu_i} \xi} \sum_{\boldsymbol{\sigma}_{T_{u_i}}: \sigma_{u_i}=-1} e^{-H_{T_{u_i}}(\boldsymbol{\sigma}_{T_{u_i}})} \bigg)\\
     &=\, \frac{e^{h_w\xi}}{Z_{T_w}}\, \prod_{i=1}^{K_w} \;Z_{T_{u_i}} \big( e^{\beta_{wu_i} \xi}\,S_{u_i, T_{u_i}}(1) \,+\, e^{-\beta_{wu_i} \xi}\, S_{u_i,T_{u_i}}(-1) \big) \\
     &=\, \frac{\prod_{i=1}^{K_w}Z_{T_{u_i}}} {Z_{T_w}}\; e^{h_w\xi}\,\prod_{i=1}^{K_w} \;  \big( e^{\beta_{wu_i} \xi}\,S_{u_i, T_{u_i}}(1) \,+\, e^{-\beta_{wu_i} \xi}\, S_{u_i, T_{u_i}}(-1) \big)
    \end{split} \end{equation}
    for $\xi=\pm1$. Let us take the ratio,
    \begin{equation}
    \frac{S_{w,T_w}(1)}{S_{w,T_w}(-1)}  = \frac{e^{ h_w} \prod_{i=1}^{K_w} \big( e^{\beta_{wu_i}}\,S_{u_i,T_{u_i}}(1) \,+\, e^{-\beta_{wu_i}}\,S_{u_i,T_{u_i}}(-1) \big)}{ e^{-h_w} \prod_{i=1}^{K_w}  \big( e^{-\beta_{wu_i} }\,S_{u_i,T_{u_i}}(1) \,+\, e^{\beta_{wu_i}}\,S_{u_i, T_{u_i}}(-1) \big) }\;.
    \end{equation}
    Notice that $S_{w,T_w}(\pm1) \,=\, \frac{1\pm{m}_w}{2}$ and $\frac{S_{w,T_w}(1)}{S_{w,T_w}(-1)} = \frac{1+{m}_w}{1-{m}_w}$, hence
    \begin{equation} \begin{split}
    \frac{1+{m}_w}{1-{m}_w} \,&=\, e^{2 h_w}\,\prod_{i=1}^{K_w}\,\frac{ e^{\beta_{wu_i}} (1+m_{u_i}) \,+\, e^{-\beta_{wu_i}} (1-m_{u_i}) }{e^{-\beta_{wu_i}} (1+m_{u_i}) \,+\, e^{\beta_{wu_i}} (1-m_{u_i}) } \\
    &=\, e^{2 h_w}\,\prod_{i=1}^{K_w}\,\frac{ 1+\tanh(\beta_{wu_i})\,m_{u_i}}{1-\tanh(\beta_{wu_i})\,m_{u_i}}
    \;.
    \end{split} \end{equation}
    Recall that $\frac{1}{2}\log\frac{1+x}{1-x}=\tanh^{-1}(x)$. Then applying the logarithm on both sides of the previous identity, we obtain
    \begin{equation}
        \tanh^{-1}(m_w) \,=\, h_w \,+\, \sum_{i=1}^{K_w}\,\tanh^{-1}\big(\tanh(\beta_{wu_i})\,m_{u_i}\big)
    \end{equation}
    completing the proof.
\end{proof}

\begin{proof}[Proof of Lemma \ref{lemma general two points correlation}]
    The Hamiltonian splits in three contributions:
    \begin{equation}
    H_T(\boldsymbol{\sigma}) \,=\, H_{T_{w}^{(w')}}\big(\boldsymbol{\sigma}_{T_{w}^{(w')}}\big) \,+\, H_{T_{w'}^{(w)}}\big(\boldsymbol{\sigma}_{T_{w'}^{(w)}}\big) \,- \beta_{ww'}\sigma_w \sigma_{w'}
    \end{equation}
    for $\boldsymbol{\sigma}\in\{-1,1\}^{V(T)}$.
    Then, setting $S_{ww',T}(\pm1) := \mu_T(\sigma_w\sigma_{w'}=\pm1)$ and $S_{w,T_w^{(w')}}(\pm1) := \mu_{T_w^{(w')}}(\sigma_w=\pm1)$, we have: 
    \begin{equation}
        \begin{split}  
            S_{ww',T}(1) &=\, \sum_{\boldsymbol{\sigma}:\sigma_w\sigma_{w'}=1}\, \frac{e^{ -H_{T_{w}^{(w')}} \,-\, H_{T_{w'}^{(w)}} \,+\, \beta_{ww'}\sigma_w\sigma_{w'}}}{Z_T} \\[2pt]
            & =\, \frac{e^{\beta_{ww'}}}{Z_T}\, \bigg( \sum_{\boldsymbol{\sigma}: \sigma_w=1,\,\sigma_{w'}=1} e^{-H_{T_{w(w')}}-H_{T_{w'(w)}}} \,+ \sum_{\boldsymbol{\sigma}: \sigma_w=-1,\,\sigma_{w'}=-1} e^{-H_{T_{w}^{(w')}}-H_{T_{w'}^{(w)}}} \bigg) \\[2pt]
            &=\, e^{\beta_{ww'}} \frac{Z_{T_{w}^{(w')}}Z_{T_{w'}^{(w)}}}{Z_T}\, \Big(S_{w,T_{w}^{(w')}}(1)\,S_{w',T_{w'}^{(w)}}(1)\,+\,S_{w,T_{w}^{(w')}}(-1)\,S_{w',T_{w'}^{(w)}}(-1)\Big)\,.
        \end{split}
        \end{equation}
        Similarly:
        \begin{equation}
        S_{ww',T}(-1) \,=\, e^{-\beta_{ww'}} \frac{Z_{T_{w}^{(w')}}Z_{T_{w'}^{(w)}}}{Z_T}\,\Big(S_{w,T_{w}^{(w')}}(1)\,S_{w',T_{w'}^{(w)}}(-1)\,+\,S_{w,T_{w}^{(w')}}(-1)\,S_{w',T_{w}^{(w')}}(1)\Big) \,.
        \end{equation}
        Taking the ratio we find:
        \begin{equation}
        \begin{split}
            \frac{1+ \langle\sigma_w\sigma_{w'}\rangle_T}{1- \langle\sigma_w\sigma_{w'}\rangle_T}\,&=\,  \frac{S_{ww',T}(1)}{S_{ww',T}(-1)} \\
            &=\, e^{2\beta_{ww'}}\, \frac{(1+m_w^{(w')}) (1+m_{w'}^{(w)}) + (1-m_w^{(w')})  (1-m_{w'}^{(w)}) }{(1+m_w^{(w')}) (1-m_{w'}^{(w)}) + (1-m_w^{(w')})  (1+m_{w'}^{(w)})} \\
            &=\, e^{2\beta_{ww'}}\,\frac{1+m_w^{(w')}m_{w'}^{(w)}}{1-m_w^{(w')}m_{w'}^{(w)}}\,. 
        \label{expre_two_point}
        \end{split}
        \end{equation}      
        By applying the logarithm at both side and inverting, we get \eqref{eq general two points}.        
\end{proof}

\begin{proof}[Proof of Lemma \ref{lemma topology h null}]
    Let us consider the graph $\mathcal G$ with vertex set $\{1,\dots,n\}$ and weighted adjacency matrix $\mathbf{k}$. 
    Let $W(\mathcal G)$ be the set of finite walks on $\mathcal G$ (we denote a walk of lenght $s\in\mathbb N$ as $c_0c_1\dots c_s\in\{1,\dots,n\}^{s+1}$) such that
    \begin{itemize}
        \item[i)] the term $\dots cd\dots$ is allowed in a walk only if $k_{cd}\geq1$ (including the case $c=d$);
        \item[ii)] the term $\dots dcd\dots$ is allowed in a walk only if  $k_{cd}\geq2$.
    \end{itemize}
    For brevity we call *walk any walk that belongs to $W(\mathcal G)$. In other words a *walk is a walk on the graph $\mathcal G$ that may have self-edges and cannot backtrack on edges of weight $1$.
    We denote by $W_a$ the set of *walks starting from $a$, and by $W_{a}^{(b)}$ the set of *walks starting from $a$ satisfying also:
    \begin{itemize}
        \item[iii)] the initial term $ab\dots$ is allowed in a walk only if $k_{ab}\geq2$.
    \end{itemize}
    It takes a moment to understand that the classes of vertices appearing in the trees $T_a\equiv\bigcup_{s\in\mathbb N}T_a(s)$, $T_a^{(b)}\equiv\bigcup_{s\in\mathbb N}T_a^{(b)}(s)$ are exactly those appearing in the collection of *walks $W_a$, $W_{a}^{(b)}$ respectively (see Definitions \ref{kregulartree}, \ref{kregularsubtree}). In other terms
    \begin{equation} \label{eq:path-tree equivalence}
        \mathcal C\big(T_a\big) \,=\, \mathcal C\big(W_a\big) \,,\quad  \mathcal C\big(T_a^{(b)}\big) \,=\, \mathcal C\big(W_{a}^{(b)}\big) 
    \end{equation}
    where $\mathcal C(A)\subseteq\{1,\dots,n\}$ denotes the set of classes of vertices appearing in $A$.
    
    %
    Now, since the matrix $\mathbf{k}$ is irreducible, the graph $\mathcal G$ is strongly connected. 
    Moreover, *walks can visit the same vertices as standard walks: the non-backtracking condition ii) is not restrictive. Indeed a non-backtraking walk (where $\dots dcd\dots$ is never allowed) could only be stuck in a leaf $c$, that is a vertex of the graph $\mathcal G$ such that $\exists\,!\,d\neq c:\,k_{cd}\geq1$. But the hypothesis $2\leq k_c=k_{cc}+k_{cd}$ guarantees that this is not the case for *walks, since either the term $\dots dcd\dots$ or the term $\dots dccd\dots$ is allowed.
    Therefore for every $f=1,\dots,n$ there exists a *walk of type $a\dots f\,$, that is
    \begin{equation} \label{eq:Ca}
      \mathcal C\big(W_a\big)  \,=\, \{1,\dots,n\} \,.
    \end{equation}
    Now, we claim that also
    \begin{equation} \label{eq:Cab}
      \mathcal C\big(W_{a}^{(b)}\big)  \,=\, \{1,\dots,n\} \,,
    \end{equation}
    namely for every $f=1,\dots,n$ there exists a *walk of type $ba\dots f\,$.
    In order to prove it, let us distinguish three cases:
    \begin{itemize}
        \item if $k_{ab}\geq2$, then $W_{a}^{(b)}=W_a$ and \eqref{eq:Cab} follows from \eqref{eq:Ca};
        \item if $k_{ab}=1$ and removing the edge $(a,b)$ the graph $\mathcal G\smallsetminus\{(a,b)\}$ remains strongly connnected, then we have $\eqref{eq:Cab}$ (as before, since condition ii) does not restrict the set of visited vertices);
        \item if $k_{ab}=1$ and removing the edge $(a,b)$ the graph $\mathcal G\smallsetminus\{(a,b)\}$ is not strongly connnected, we denote by $\mathcal G_a$, $\mathcal G_b$ its two strongly connected components containing $a,b$ respectively.
        Since $k_a\geq2$, there is a *walk $a\dots c a$ of positive lenght in $\mathcal G_a$ that starts from $a$ and comes back to $a$ after visiting all the vertices of $\mathcal G_a$; in particular $c\neq b$. Then:
        \begin{equation} \label{eq:Pab}
            W_{a}^{(b)} \,\supseteq\, \{a\dots ca\}\times W_{b}^{(a)}\,.
        \end{equation}
        Now, there is a *walk $b\dots d$ in $\mathcal G_b$ that starts from $b$ and visits all the vertices of $\mathcal G_b$. Thus $W_{b}^{(a)}\supseteq\{b\dots d\}$. Therefore plugging into \eqref{eq:Pab} we get
        \begin{equation}
            W_{a}^{(b)} \supseteq \{a\dots cab\dots d\} \quad\Rightarrow\quad \mathcal C\big(W_{a}^{(b)}\big)\,=\,\{1,\dots,n\} \;.
        \end{equation}
    \end{itemize}
    Identity \eqref{eq:Cab} together with \eqref{eq:path-tree equivalence} conclude the proof of the lemma.
\end{proof}

\begin{remark}
    $k_1,\dots,k_n\geq 2$ is a necessary and sufficient condition for $\mathcal C\big(W_{a}^{(b)}\big)=\{1,\ldots,n\}$ for all $(a,b)\in \mathcal E$. However, if some $k_a=1$ Corollary \ref{corollary_final_allnonnegative_fields_one_positive} can be extended, noticing that in this case $\bar{z}_a^{(b)}= h_a$ for $b\in\mathcal E_a$, and verifying case-by-case whether for every $(c,d)\in\mathcal E\smallsetminus\{(a,b)\}$ there exists $f\in \mathcal C\big(W_{c}^{(d)}\big)$ such that $h_f>0\,$.
\end{remark}

\begin{definition}\label{def: non-cyclic }   
    Consider the graph $\mathcal G$. We say that a couple of vertices $(a,b)$ is \textit{+adjacent} if $a-(b-1)\equiv 0 \; (\textrm{mod}\;n)$; we sayt that $(a,b)$ is \textit{-adjacent} if  $a-(b+1)\equiv 0 \; (\textrm{mod}\;n)$. $\mathcal G$ is called \textit{simply cyclic} if
    \begin{equation}
        k_{ab}\,>0  \quad\Leftrightarrow\quad (a,b) \textrm{ is +adjacent or -adjacent}
    \end{equation}
    and $\nexists$ $a,b,c,d\in \{1, . . . , n\}$ such that $ab$ \textit{+adjacent}, $cd$ \textit{-adjacent}, $k_{ab}\geq 2$, $k_{cd}\geq 2$.
\end{definition}

\begin{proof}[Proof of Lemma \ref{lemma: irreducibility of M}]
    Let us introduce the matrix $\boldsymbol {\tilde M}$ indexed by $\mathcal E\times\mathcal E$ with entries
    \begin{equation}
    \tilde{M}_{(a,b),(c,d)} \,:=\, \begin{cases}
        \,1 &\textrm{\,if } (a=d,\,b=c,\, k_{ac}\geq 2) \textrm{ or } (a=d,\,b\neq c) \\
        \, 0 &\textrm{\,if }a\neq d  \textrm{ or }  (a=d,\,b=c,\, k_{ac}=1)
    \end{cases} 
    \end{equation}
    for every $(a,b)\in \mathcal E$, $(c,d)\in \mathcal E$. Since $\boldsymbol{\tilde M}$ and $\boldsymbol{M}$ have the same positive entries, $\boldsymbol{\tilde M}$ is irreducible if and only if  $\boldsymbol{M}$ is.
    
    Notice that for $s\geq1$ the element $\big(\boldsymbol{\tilde{M}}^s\big)_{(a,b),(c,d)}$ represents the number of possible *walks on the graph $\mathcal G$  (as defined in the Proof of Lemma \ref{lemma topology h null}) of type
    $ba\ldots dc$
    and length $s+1$. Therefore, in order to prove that the matrix $\boldsymbol{\tilde M}$ is irreducible, we have to show that for every $(a,b)\in\mathcal E,\,(c,d)\in\mathcal E$ there exists a *walk on $\mathcal G$ of type
    \begin{equation}
        ba\ldots dc \,.
    \label{toprove walk}
    \end{equation}
    In Lemma \ref{lemma topology h null} we have already proved that for every $(a,b)\in \mathcal E,\;d\in\{1,\ldots,n\}$ there exists a *walk of type
    \begin{equation}
        ba\ldots d  \;.
    \label{proved walk}
    \end{equation}
    Let us write the latter *walk as $ba\ldots ed\,$.
    If $k_{dc}\neq 1$ or $e\neq c$, the existence of \eqref{toprove walk} follows immediately since a possible *walk is $ba\ldots edc$.
    Suppose now $k_{dc}=1$ and $e=c$, since $k_d\geq2$ there exists a *walk $ba\ldots cdf$ for some $f\neq c$. To continue from $df$ we have the following alternatives:
    \begin{itemize}
        \item if $k_{fd}\geq 2$, then we can backtrack and $ba\ldots cdfdc$ is a possible *walk;
        \item if $k_{fd}=1$, since $k_f\geq2$ and $\mathcal G$ is strongly connected and not a simple cycle, there exists a *walk of type $dfg\ldots hf$ with $g,h\neq d$. Then $hfdc$ is a possible *walk. Gluing them together we find the desired *walk: $ba\ldots cdfg\ldots hfdc$.
    \end{itemize}
\end{proof}

\section{Correlation Inequalities}
In this second Appendix we report the classical GKS (Griffiths-Kelly-Sherman) \cite{griffiths1967correlations, KS}, FKG (Fortuin-Kasteleyn-Ginibre) \cite{fortuin1971correlation} and GHS (Griffiths-Hurst-Sherman) \cite{griffiths1970concavity,lebowitz1974ghs} correlation inequalities, which are involved in many proofs of the present paper.

We consider an Ising spin system of $N$ binary spins, $\sigma_i=\{-1,+1\}$ for $i= 1,...,N$, on a general graph $G_N(V,E)$, $|V|=N$, with Hamiltonian 
\begin{equation}
    H(\boldsymbol{\sigma}) = -\sum_{ij\in E}\beta_{ij}\sigma_i\sigma_j - \sum_{i\in V}h_i\sigma_i
\end{equation}
Under the Gibbs (Boltzmann) measure 
\begin{equation} 
    \mu_I (\boldsymbol{\sigma}) = \frac{e^{-H(\boldsymbol{\sigma})}}{\sum_{\boldsymbol{\sigma}} e^{-H(\boldsymbol{\sigma})}}
\end{equation}
so expressing the average of an observable (function of the spins) as
\begin{equation} 
    \langle f(\boldsymbol{\sigma}) \rangle = \sum_{\boldsymbol{\sigma}}\frac{f(\boldsymbol{\sigma}) e^{-H(\boldsymbol{\sigma})}}{\sum_{\boldsymbol{\sigma}} e^{-H(\boldsymbol{\sigma})}}
\end{equation}
the following inequalities, under the corresponding assumptions on the Hamiltonian, hold. Before listing the inequalities, let us also recall that defining $m_i = \langle \sigma_i \rangle$ we have that
$\frac{\partial \;m_i}{\partial\;h_j} =$\small$ \langle \sigma_i \sigma_j\rangle -  \langle\sigma_i\rangle\langle \sigma_j\rangle$ and\\
$\frac{\partial^2 \;m_i}{\partial\;h_j\;\partial\;h_k} = $\small$\langle\sigma_i \sigma_j \sigma_k\rangle+ 2\langle\sigma_i\rangle\langle \sigma_j\rangle\langle \sigma_k\rangle - (\langle\sigma_i\rangle\langle \sigma_j \sigma_k\rangle+ \langle\sigma_j\rangle\langle\sigma_i \sigma_k\rangle + \langle\sigma_k\rangle\langle \sigma_i \sigma_j\rangle )$.

\begin{theorem}[GKS inequalities]\label{GKS}
    If $h_i\geq0 \;\forall \;i$, $\;\;\beta_{ij}\geq0 \;\forall\;i,j\;\;$    then
    \begin{equation}
        \langle\sigma_A \sigma_B\rangle \;\geq\; \langle\sigma_A\rangle\langle \sigma_B\rangle
    \end{equation}
    where $A,B\subseteq V$ and $\sigma_{A} = \prod_{i\in A} \sigma_i$.
\end{theorem}
\begin{remark} \label{GKS_rem}
    Under the conditions of the GKS, $\forall i,j,k,l$
    \begin{align}
        &\frac{\partial \;m_i}{\partial\;h_j} = \langle\sigma_i \sigma_j\rangle-  \langle\sigma_i\rangle\langle \sigma_j\rangle \;\geq 0  \label{GKS_i_h}\\
        &\frac{\partial \;m_i}{\partial\;\beta_{jk}} = \langle\sigma_i \sigma_j \sigma_k\rangle -  \langle\sigma_i\rangle\langle\sigma_j\sigma_k\rangle \;\geq 0 \label{GKS_i_beta} \\
        & \frac{\partial \;\langle \sigma_i \sigma_j\rangle}{\partial\;h_k} = \langle\sigma_i \sigma_j \sigma_k \rangle-  \langle\sigma_i\sigma_j\rangle\langle \sigma_k\rangle \;\geq 0   \label{GKS_ij_h} \\
        &\frac{\partial \;\langle \sigma_i \sigma_j\rangle}{\partial\;\beta_{kl}} = \langle\sigma_i \sigma_j \sigma_k \sigma_l\rangle -  \langle\sigma_i\sigma_j\rangle\langle\sigma_k\sigma_l\rangle \;\geq 0 \label{GKS_ij_beta}
    \end{align}
\end{remark}

\begin{theorem}[FKG inequalities] \label{FKG}
    For any external fields $h_i\in\mathbb R\;\forall i$, for $\beta_{ij}\geq0 \;\forall\;i,j\;\;$, if $f(\vec{\sigma})$ and $g(\vec{\sigma})$ are increasing functions of the spins, in the sense that $f(\vec{\sigma}) := f(\sigma_1,...\sigma_i,...,\sigma_N)$  is such that
    \begin{equation}
        f(\sigma_1,...\sigma_i=-1,...,\sigma_N) \leq f(\sigma_1,...\sigma_i=1,...,\sigma_N)
    \end{equation}
    for any $i$, and same for $g(\vec{\sigma})$; then 
    \begin{equation} 
        \langle fg\rangle\;\geq\;\langle f\rangle\langle g\rangle
    \end{equation}
\end{theorem}
\begin{remark}\label{FKG_field}
    If $\beta_{ij}\geq0, \;\forall\;i,j\;\;$
    \begin{equation} 
        \frac{\partial \;m_i}{\partial\;h_j} = \langle\sigma_i \sigma_j\rangle -  \langle\sigma_i\rangle\langle \sigma_j\rangle \;\geq 0
    \end{equation}
    (since we apply FKG inequality choosing $f(\vec{\sigma}) = \sigma_i$ and $g(\vec{\sigma}) = \sigma_j$, both increasing functions).
\end{remark}

\begin{theorem}[GHS inequalities]\label{GHS}
    If $h_i\geq0 \;\forall \;i$, $\;\;\beta_{ij}\geq0 \;\forall\;i,j\;\;$    then
    \begin{equation} 
        \frac{\partial^2 \;m_i}{\partial\;h_j\;\partial\;h_k}\leq 0
    \end{equation}
    $\forall \; i,j,k$ not necessairly distinct.
\end{theorem}

\begin{proof}[Proof of  Remark \ref{remark_proof_GKS}]\label{proof_GKS_existence}
    Consider the following $\mathbf{k}$-regular trees:
    \begin{itemize}
        \item $T_1$ of depth $t$ and positive boundary conditions obtained by fixing the fields of the spins at the boundary to $+\infty$
        \item $T_2$ of depth $t+1$, with positive infinite fields not only at the last generation, but also at the generation $t$
        \item $T_3$ of depth $t+1$, with positive infinite fields at the last generation
    \end{itemize}
    For any $(a,b)\in \mathcal E$, 
    \begin{equation}
    \begin{split}
        &z_a^{(b)}(s-1,+\infty) \equiv z_a^{(b)}(s-1,T_1) = z_a^{(b)}(s,T_2)    \quad \quad    s\geq 1 \\
        &z_a^{(b)}(s,+\infty) \equiv z_a^{(b)}(s,T_1) = z_a^{(b)}(s,T_3)    \quad \quad  \;\;\;\;\;\;\;\;\;\;\;\;  s\geq 0
    \end{split}
    \label{eq equivalences T1 T2 T3 h_big}
    \end{equation}
    where the boundary (and second-last generation) conditions are specified through $T_{1/2/3}$.\\
    For transforming $T_2$ into $T_3$ we have to remove the fields at the $t$-th generation, thus the GKS inequalities \eqref{GKS_i_h} hold and
    \begin{equation}
        z_a^{(b)}(s,T_2) \geq z_a^{(b)}(s,T_3)     \quad \quad    s\geq 0
    \end{equation}
    Thus, inserting \eqref{eq equivalences T1 T2 T3 h_big}, we obtain
    \begin{equation}
        z_a^{(b)}(s-1,+\infty) \geq z_a^{(b)}(s,+\infty)   \quad \quad    s\geq 1
    \end{equation} 
    which proves the monotonic non-increasing behaviour of the recursion for $h_*=+\infty$.\\
    \\   
    For free boundary conditions ($h_*=0$), we consider the following $\mathbf{k}$-regular trees:
    \begin{itemize}
        \item $T_1$ of depth $t$ and all free boundary conditions (in the sense that there is no added field at the boundary, $h_*=0$)
        \item $T_2$ of depth $t+1$, with no added fields at the boundary and no edges among the spins of the last and second-last generations
        \item $T_3$ of depth $t+1$, with no added field at the boundary
    \end{itemize}
    Analogously, for any $(a,b)\in \mathcal E$, 
    \begin{equation}
    \begin{split}
        &z_a^{(b)}(s-1,0) \equiv z_a^{(b)}(s-1,T_1) = z_a^{(b)}(s,T_2)    \quad \quad    s\geq 1 \\
        &z_a^{(b)}(s,0) \equiv z_a^{(b)}(s,T_1) = z_a^{(b)}(s,T_3)    \quad \quad    \;\;\;\;\;\;\;\;\;\;\;\;  s\geq 0
    \end{split}
    \label{eq equivalences T1 T2 T3 h_small}
    \end{equation}
    For transforming $T_2$ into $T_3$ we have to add to $T_2$ the the missing ferromagnetic couplings between the last two generations. As we only increase the weight of some edges, the GKS inequalities \eqref{GKS_i_beta} hold and 
    \begin{equation}
        z_a^{(b)}(s,T_2) \leq z_a^{(b)}(s,T_3)     \quad \quad    s\geq 0
    \end{equation}
    Thus, inserting \eqref{eq equivalences T1 T2 T3 h_small}, we obtain
    \begin{equation}
        z_a^{(b)}(s-1,0) \leq z_a^{(b)}(s,0)   \quad \quad    s\geq 1
    \end{equation} 
    which proves the monotonic non-decreasing behaviour of the recursion for $h_*=0$.\\
\end{proof}

\begin{proof}[Proof of Remark \ref{remark_concavity_GHS}]
    By applying the chain rule, we obtain
    \begin{equation}
    \begin{split}
        \frac{d^2 z_a^{(b)}(s,h_*)}{dh_*^2} &= \frac{d}{dh_*}  \sum_{l = 1,...,n} \frac{\partial z_a^{(b)}(s,h_*)}{\partial h'_{*l}} =  \sum_{l,p = 1,...,n} \frac{\partial^2 z_a^{(b)}(s,h_*)}{\partial h'_{*p} \partial h'_{*l}} \leq 0
    \label{GHS_POSITIVE_concavity_cond}
    \end{split}
    \end{equation}
    where $h'_{*l} = h_{l} + h_*$. The last inequality follows from the GHS theorem (\ref{GHS}), as the fields and couplings are all non-negative. 
\end{proof}

\section*{Acknowledgements}

PC and EM was supported by the EU H2020 ICT48 project Humane AI Net contract number 952026; by the Italian Extended Partnership PE01 - FAIR Future Artificial Intelligence Research - Proposal code PE00000013 under the MUR National Recovery and Resilience Plan; by the project 
PRIN 2022 -  Proposal code: J53D23003690006.  
The work is partially supported by GNFM (Indam), EU H2020 ICT48 project Humane AI Net. The authors are affiliated to GNFM INdAM. 

\printbibliography

\end{document}